\documentclass[useAMS,usenatbib]{mn2e}

\usepackage{amssymb}
\usepackage{aas_macros}
\usepackage{natbib}
\usepackage{times}
\usepackage{graphicx}
\usepackage{url}

\newcommand{\gadget}{{\sc GADGET}}
\newcommand{\flash}{{\sc FLASH}}
\newcommand{\galform}{{\sc GALFORM}}

\newcommand{\ltsima}{\hbox{$\; \buildrel < \over \sim \;$}}

\newcommand{\muEst}{1.15}
\newcommand{\nuEst}{0.16}
\newcommand{\betaEst}{13}

\title[Supernova driven winds]{How supernova explosions power galactic winds}
\author[Creasey, Theuns \& Bower]  {\parbox[h]{160mm} { 
    Peter Creasey$^{1}$\thanks{E-mail: p.e.creasey@durham.ac.uk},
    Tom Theuns$^{1,2}$ and
Richard G. Bower$^1$}
  \vspace{6pt}\\
  $^1$Institute for Computational Cosmology, Department of Physics,
  University of Durham, South Road, Durham, DH1 3LE, UK\\
  $^2$Department of Physics, University of Antwerp, Campus Groenenborger, Groenenborgerlaan 171, B-2020 Antwerp,  Belgium}

\begin{document}

\date{\today}
\pagerange{\pageref{firstpage}--\pageref{lastpage}} \pubyear{2010}

\maketitle

\label{firstpage}

\begin{abstract}
Feedback from supernovae is an essential aspect of galaxy formation. In order to improve subgrid models of feedback we perform a series of numerical experiments to investigate how supernova explosions shape the interstellar medium in a disk galaxy and power a galactic wind. We use the \flash\ hydrodynamic code to model a simplified ISM, including gravity, hydrodynamics, radiative cooling above $10^4$~K, and star formation that reproduces the Kennicutt-Schmidt relation. By simulating a small patch of the ISM in a tall box perpendicular to the disk, we obtain sub-parsec resolution allowing us to resolve individual supernova events.
The hot interiors of supernova explosions combine into larger bubbles that sweep-up the initially hydrostatic ISM into a dense, warm cloudy medium, enveloped by a much hotter and tenuous medium, all phases in near pressure equilibrium. The unbound hot phase develops into an outflow with wind speed increasing with distance as it accelerates from the disk.  We follow the launch region of the galactic wind, where hot gas entrains and ablates warm ISM clouds leading to significantly increased mass loading of the flow, although we do not follow this material as it interacts with the galactic halo.

We run a large grid of simulations in which we vary gas surface density, gas fraction, and star formation rate in order to investigate the dependencies of the mass loading, $\beta \equiv \dot M_{\rm wind}/\dot M_\star$. In the cases with the most effective outflows we observe a $\beta$ of 4, however in other cases we find $\beta \ll 1$. We find that outflows are more efficient in disks with lower surface densities or gas fractions. 
A simple model in which the warm cloudy medium is the barrier that limits the expansion of the blast wave reproduces the scaling of outflow properties with disk parameters at high star formation rates.
We extend the scaling relations derived from an ISM patch to infer an effective mass loading for a galaxy with an exponential disk, finding that the mass loading depends on circular velocity
as $\beta \propto V_{\rm d}^{-\alpha}$ with $\alpha \approx 2.5$ for a model which fits the Tully-Fisher relation. Such a scaling is often assumed in phenomenological models of galactic winds in order to reproduce the flat faint end slope of the mass function. Our normalisation is in approximate agreement with observed estimates of the mass loading for the Milky Way. The scaling we find sets the investigation of galaxy winds on a new footing, providing a physically motivated sub-grid description of winds that can be implemented in cosmological hydrodynamic simulations and phenomenological models.
\end{abstract}
\begin{keywords}
hydrodynamics, galaxies: formation, methods: numerical, galaxies: ISM, 
\end{keywords}


\section{Introduction}
Feedback is an essential aspect of galaxy formation models. It is invoked to suppress the formation of 
large numbers of small galaxies \citep{Rees_1977, WhiteRees_1978, White_1991}. While photo heating can suppress star formation in the smallest halos, it cannot explain the low efficiency of SF in halos more massive than $10^9\,\rm M_\odot$ \citep{Efstathiou_1992, Okamoto_2008}. Feedback is also invoked to explain why such a small fraction of the baryons are in stars today \citep{Fukugita_1998, Balogh_2001}. An efficient feedback implementation also appears essential for simulations to produce realistic looking disk galaxies \citep{Scannapieco_2011, McCarthy_2012}.  Observations of galactic winds at low \citep{Heckman_1990, Heckman_2000} and high $z\sim 3$ redshift \citep{Pettini_2001} do show gas with a range of temperature and densities moving with large velocities of 100s of km~s$^{-1}$ with respect to the galaxy's stars, although the interpretation in terms of mass loss is complicated by the multi-phase nature of the wind (see e.g. \citealp{Veilleux_2005} for a recent review). Complimentary evidence for outflows comes from the high metal abundance detected in the IGM \citep{Cowie_1995}, even at low densities \citep{Schaye_2003, Aguirre_2004}. Numerical simulations make it plausible that galactic winds are responsible for this metal enrichment \citep{Cen_1999, Aguirre_2001, Theuns_2002, Aguirre_2005, Oppenheimer_2006, Tescari_2011}, with low-mass galaxies dominating the enrichment of the bulk of the IGM \citep{Booth_2012}.

The sheer amount of energy released by supernovae (SNe) make the injection of energy into the interstellar medium (ISM) by SN explosions a prime candidate for driving galactic winds \citep{Dekel_1986}. However it is challenging to understand in detail how SNe regulate the transfer of mass and energy between the different phases of the ISM, as envisaged in the model of \citet{McKee_Ostricker_1977}, and how and when this leads to the emergence of a galactic wind. \citet{Efstathiou_2000} and \citet{Silk_2001} extend the \citet{McKee_Ostricker_1977} model to examine how such interactions lead to self-regulation of star formation. They show that the properties of the galactic wind can be broadly understood once a temperature and density for the hot phase is found. This requires a model of evaporation of cold and warm clouds, yet without a more detailed understanding of the geometry and turbulence, we can go little further than steady spherically symmetric conduction models, which go back to \citet{Cowie_McKee_1977}. Even if feedback is indeed due to SNe, it is not yet clear whether this is a consequence of their injection of hot gas, of turbulence \citep{MacLowKlessen_2004, Scannapieco_2010}, of cosmic rays \citep{Jubelgas_2008}, of the combined effects of magnetic fields, cosmic rays, and the galaxy's differential rotation \citep{Kulpa_2011}, or all of the above.

Full hydrodynamic modelling of the interplay between the various components of the ISM in a Milky Way-like galaxy in a proper cosmological context is not yet currently possible due to the large range of scales involved, with density ranging from $4\times 10^{-31}$~g~cm$^{-3}$ outside of halos to $\sim 10^{-20}$~g~cm$^{-3}$ in cold clouds, temperatures from a few Kelvin inside star forming regions to $\sim 10^8 \rm K$ inside SN remnants, and time scales from a few thousand year for the propagation of a SN blast wave inside the ISM to $\sim 10^{10}$~years for the age of the Galaxy. Excitingly, such full hydro-dynamical modelling begins to be possible for higher redshift dwarfs (e.g \citealp{Wise2008}), but for the moment models of larger galaxies at $z\sim 0$ are limited to simulating a patch of galactic disk. In addition, we would also like to identify and understand the physics that is important in driving material from the galactic disk, and so it is desirable to have a series of numerical experiments. This is the approach we will follow in this paper.

We begin by discussing constraints on galactic winds derived from current theoretical models of galaxy formation, and place our work in the context of comparable approaches. In section \ref{sect:simulations} we introduce the set-up of our own simulations. Briefly, we use a very simple model of the ISM which neglects the cold phase, and which is stirred by hot gas injected by SN explosions. Next we demonstrate that our sub-pc simulations have sufficient resolution to resolve individual explosions, and illustrate the behaviour of both the ISM and of the wind for a reference model with properties chosen to be similar to that of the solar neighbourhood. In Section~\ref{sect:statistics} we vary the properties of the simulated ISM (total and gas surface densities, star formation rate, cooling rate), and investigate if and when a wind is launched, and how its properties depend on that of the ISM. We obtain scaling relations of the wind to the ISM and
apply them in Section~\ref{sect:evolution} to predict wind properties for a full galactic disk, and investigate how the wind properties depend on the galaxy properties. We summarise in Section~\ref{sect:conclusions}.

\section{Constraints on galactic winds}\label{sec:WindConstraints}
\subsection{Model requirements and observations}
\label{sect:mass_loading}
We will assume that the baryon fraction in Milky-Way-sized halos, and halos of lower mass, falls significantly below the cosmological value, $f_b = \Omega_b / \Omega_M$ due to the action of a galactic wind. Let the gaseous mass outflow rate from this wind be $\dot{M}_{\rm wind}$, and the star formation rate $\dot{M}_\star$. A simple way to parameterise the efficiency of the SN-driven wind in removing baryons from the halo, is its {\em mass loading}, i.e. the ratio
\begin{equation}\label{eq:beta}
\hat \beta \equiv \frac{\dot{M}_{\rm wind}}{\dot{M}_\star}\,,
\end{equation}
where our $\hat \beta$  is equivalent to the $\beta$ of \cite{Stringer_2011}. We introduce the hat in order to distinguish the average mass loading for a galaxy, $\hat \beta$, from a local mass loading $\beta$ at some point on the disk. If a galaxy exhausts its gas supply in star formation (and does not recycle wind material) then we will be left with a gas poor galaxy with baryon fraction reduced by a factor $1 / (1+\hat \beta)$.

In order to infer the fraction of baryons ejected from galaxies we can use the statistics of galaxies and dark matter halos. The number density of halos as a function of their mass can be approximated for masses below the exponential cut-off scale as a power law \citep{PressSchechter_1974, Reed_2007},
\begin{equation}
\frac{ {\rm d} n}{{\rm d} \log M}  \propto M^{-0.9} \,.
\end{equation}
Contrast this with the slope of the galaxy stellar mass function at low masses,
\begin{equation}
\frac{ {\rm d} n }{{\rm d} \log M_\star}  \propto M_\star^{1+\alpha} \, ,
\end{equation}
where observationally $\alpha$ is found to be in the range $\left[-1.5,-1\right]$, (see e.g. \citealp{Blanton_2003, Blanton_2005B, Baldry_2005, Baldry_2012, Li_White_2009}). Naively identifying each dark matter halo with a galaxy of a given stellar mass (e.g. \citealp{Guo_2010}) yields a galaxy mass to halo mass relation of $M_\star \propto M^{-0.9/(\alpha+1)}$. Identifying the stars as the main baryonic component implies a mass loading that scales with halo mass relatively steeply as (see also \citealp{Stringer_2011})
\begin{equation}
1+\hat \beta = f_{\rm b}\,\frac{M}{M_\star} \propto M^{(1.9+\alpha)/(1+\alpha)} \propto M^{-0.8} \, ,
\label{eq:betascale}
\end{equation}
where we substituted a faint end slope of $\alpha=-1.5$ to derive the last exponent. Notably, this exponent $\to \infty$  as $\alpha \to -1$ and falls to zero as $\alpha \to -1.9$, as such it is rather poorly constrained even by a well measured slope of the galaxy stellar mass function at low masses. One can infer not only that at low masses the mass loading $\hat \beta \gg 1$ but also that it is strongly increasing towards lower-mass galaxies.

Assume star formation results in the explosion of $\varepsilon_{100}$ supernovae per $100 \, \rm M_\odot$ of stars formed, each with energy $E_{\rm SN}$, and that a fraction $\eta_T$ gets converted into kinetic energy of an outflow. Neglecting other sources of energy then implies that
\begin{equation}
\hat \beta \,v^2_{\rm wind} = 2 (710\,{\rm km}{\rm s}^{-1})^2\,\eta_T\,\varepsilon_{100}\,{E_{\rm SN}\over 10^{51}\,{\rm erg}} \,, \label{eq:eta_T}
\end{equation}
where $v_{\rm wind}$ is the wind speed. If $\varepsilon_{100}$, the thermalisation factor $\eta_T$ and $E_{\rm SN}$ are all constants, then the product $\hat \beta v^2_{\rm wind}$ is also a constant. In this case large values of $\hat \beta$ imply lower wind speeds, and vice versa.
If the mass-loading $\hat \beta$ indeed increases with decreasing halo mass, then of course eventually $\hat \beta$ may become so large that the wind can no longer escape from the galaxy's potential well. Such small halos may be subject to other destructive mechanisms, such as evaporation by re-ionization or obliteration by the explosions of the first stars. For massive halos, in order for the wind to escape it requires high wind speeds, implying low mass loading. The semi-analytical model of galaxy formation presented recently by \cite{Bower_2011} imposes similar constraints on galactic winds to obtain fits to the faint-end of the galaxy mass function as inferred from
our naive expectations: galactic winds need to have values of the mass loading $\hat \beta\sim 1$ for Milky Way-like galaxies, with an indication that $\hat \beta$ increases even further towards lower masses. The best fitting models have $\hat \beta \sim 10$ giving $v_{\rm wind} \sim 300 \; \rm km\, s^{-1}$.

Numerical simulations of galaxy formation also try to implement galactic winds with similar properties. Cosmological simulations such as \citet{Oppenheimer_Dave_2008} essentially implement the mass loading by hand by de-coupling the winds from the surrounding gas. More advanced techniques impose some constraints during the early stages of a burst of star formation when it is beneath the simulation resolution, but later allow the gas distribution to evolve normally and the mass loading to emerge. Progress in this area has been made by simulations such as \citet{Dubois_2008} and \citet{Shen_2010}. Generally these include efficient feedback in an effort to produce a reasonable galaxy population, although they struggle to produce significant winds to remove enough baryons from Milky Way-sized galaxies. 

The OWLS simulations \citep{Schaye_2010} examined a variety of feedback prescriptions and models with efficient feedback in terms of a strong galactic winds fit a variety of  properties of the galaxy population, including the Tully-Fisher relation \citep{McCarthy_2012}. However, in such models the properties of the winds are still part of the sub-grid modelling, i.e. the wind's properties are not computed but rather are simply imposed. This is required since the mass of gas entrained by a single supernova is a tiny fraction of the mass resolution element of the simulation \citep{Creasey_2011}.

In order to directly simulate the \emph{generation} of galactic winds requires a much higher resolution than can be reached in current cosmological simulations, as the sites of energy injection must be resolved (discussed further in section \ref{sect:simulations}). In order to relax these constraints, many simulators have either moved to high redshift (where the volumes are smaller), or modified the SNe in some way (such as aggregation of the energy injection). Examples of the former include \citet{MacLow_1999, Fujita_2004, Wise2008} and \citet{Powell_2011}, all of which struggle to produce mass loadings above unity except \cite{Wise2008}, who had massive Population III progenitors for their SNe. Examples of the latter include \cite{Dubois_2008} and \cite{Hopkins_2011} with similar results, although \cite{Hopkins_2011} saw significant improvement by including the winds from massive stars.

Despite having a different focus, there are also a number of studies of a high resolution SN driven ISM which have similar set-up to the current work, although they do not investigate the properties of their winds. \citet{Joung_MacLow_2006, Joung_MacLow_2009, Hill_2012} and whilst this paper was being prepared \cite{Gent_2012} have all modelled a SN driven ISM in a column through a galactic disk, driving a vertical wind. These studies investigate the structure of the ISM, however their wind properties appear qualitatively similar to ours. On larger scales \cite{Stevens_2000} and \cite{Cooper_2008} have extended these to an approximation of the galaxy M82, although again the resolution restrictions severely limit the simulation time and SN energy injection prescription.

There are compelling theoretical reasons to expect a high mass loading in galaxy winds, but are such winds seen in practise? The observational evidence for galactic outflows, at least in {\em starburst galaxies}, is extremely strong \citep{Heckman_1990, Heckman_2000, Pettini_2001, Martin_2005, Martin_2002, Strickland_2009}. Unfortunately it is notoriously difficult to constrain the wind properties from the data directly, partly because of uncertain metallicity and ionisation corrections needed to translate between the observed ion outflow and inferred total wind values, and partly because observing the wind in the spectrum of its galaxy does not provide spatial information of where the absorbing gas is located (\citealp{Bouche_2011}, but see \citealp{Wilman_2005, Swinbank_2009} for a few cases of resolved studies of winds). The outflowing gas is likely multi-phase in nature, complicating further the interpretation of the data.
The picture for non starburst galaxies is even more complex, with \citet{Strickland_2009} noting the lack of evidence for superwinds in such galaxies. As \citet{Chen_2010} point out, however, the evidence for the high velocity outflows come from blueshifted absorbers such as Na D that are tracing cooler material which is a fraction of the wind (or MgII, for example \citealp{Weiner_2009} in the Deep2 galaxies). As far as it can be measured, the velocity of the outflow seems to be only weakly dependent on the SFR \citep{Rupke_2005b}. Probing the circum-galactic medium around galaxies with a sight line to a background quasar allowed  \cite{Bouche_2011} to infer values of $\hat \beta=2-3$ and wind speeds $v_{\rm wind}=150-300$~km~s$^{-1}$ for a set of L$_\star$ galaxies at redshift $z\sim 0.1$. They claim these wind speeds are in fact below the escape speed, and hence we may be observing a galactic fountain rather than a proper outflow. 

The picture of SNe as the driver of galactic winds also has consequences in terms of the metallicity of the galaxy. As SNe inject both metals and energy we expect and find a corresponding metallicity deficit for low mass galaxies \citep{Tremonti_2004}, interest in which goes back to \citet{Larson_1974}. Both simple models \citep{Peeples_2011, Dayal_2012} and simulations \citep{Finlator_2008} show that galactic winds are an essential ingredient to obtain the observed mass-metallicity relation in galaxies.

Summarising, observations provide strong evidence for the presence of galactic winds in star forming galaxies, but the parameters of such winds are 
currently not tightly constrained. Models that make recourse to such winds to quench star formation require relatively high values of the wind's mass loading, $\hat \beta\sim 1$ for MW-like galaxies, with $\hat \beta$ increasing for lower mass galaxies. But do SNe-driven winds indeed meet these requirements, and if they do, why?

\subsection{Resolving SNe in the ISM}\label{sec:SelfConsist}
Ideally one would wish to probe the efficiency with which star formation can drive winds with simulations that self consistently included all the relevant physics, i.e. a full galaxy containing a star forming ISM, those stars subsequently redistributing their energy as type II SNe explosions, including outflows and cosmological infall. Unfortunately the range of scales involved in this problem makes such an approach currently computationally infeasible. To progress we must either truncate our resolution at some scale before we have fully resolved the physics, or to truncate our physics such that the available resolution becomes sufficient. The former route is one where we assume that we understand the physical processes to a certain degree and make our best effort at the calculation, forcing us to go deeply in to convergence studies. The latter is that of the numerical experiment where it is assumed that a certain amount of numerical calculation is possible and we make our best effort to include the processes, requiring us to make full comparison with the real Universe to test these assumptions (many simulations, are, of course, a mixture of these approaches). Our focus will be on the latter case, that of the numerical experiment. We will also restrict ourselves to looking at the \emph{launch} region of the galactic wind, where gas is expelled from the galaxy but not necessarily from the halo. This is consistent with what is needed to improve subgrid models in semi-analytic models and hydro simulations.

The motivation for our choice of scale relates to the need to resolve individual SN blast waves as they sweep the ISM (as for example described by \citealp{Cox_72}). 
Briefly, such explosions involve three distinct stages (e.g. \citealp{Truelove_1999}), beginning with the very early stage during which SNe ejecta expand almost freely into the ISM. 
As the ejecta sweep-up ISM preceded by a shock, eventually a reverse shock will run back into the ejecta, heating them to very high temperatures, signalling the start of the Sedov-Taylor stage \citep{Sedov_59, Taylor_1950}. In both stages radiative cooling is negligible and consequently they can be described by similarity solutions, but the transition between them cannot. Finally at late times, the hot interior of shocked ejecta cools radiatively,  and the swept-up shell of ISM and ejecta continue to \lq snow-plough\rq\ further into the ISM, conserving momentum. \cite{Thornton_1998} examine these last two states using a set of 1 dimensional simulations of the evolution of explosions in a uniform ISM, examining in detail the transition from the ST-phase to the snow-plough phase. They claim that radiative cooling is efficient enough that typically only 10 per cent of the initial blast energy is transferred to the ISM. Notably, the amount of gas heated by these explosions is not a linear function of the SN energy, indeed it is sub-linear, and thus we may expect that aggregating the energy injection of many SNe into a single event will underestimate the amount of gas heated.

We would in principle like to resolve the earliest phase of the explosions when ejecta dominate, but in this paper we restrict ourselves to initiate our SNe in the Sedov-Taylor phase. The transition between ejecta-dominated and ST-phase occurs approximately when the shock has swept-up an amount of 
of ISM mass that is comparable to that originally ejected. In low density regions the size of the bubble where the transition happens may then be relatively large, and it would be worthwhile investigating whether this matters; we intend to do so in future work.  Given this limitation, and for the typical ISM densities near the centre of the disk in our simulations, it then suffices to resolve scales of order of a few parsecs to fully capture the cooling of the swept-up shell of ISM (e.g. \citealp{Cox_72}), and such a simulation will be able to resolve both the cooling and some part of the adiabatic phase of the remnant. 

As such the dependence of our question upon sub-parsec phenomena can be seen only in two key areas, raising the following questions
\begin{enumerate}
\item Star formation occurs on these scales, and thus controls the distribution (in time and space) of supernovae. Does this affect the properties of the galactic wind, for example because supernovae explode in high density environments and/or near to other supernovae?
\item The medium that the SNe drive into contains structures on sub-parsec scales, for example cores of molecular clouds. Does this departure from a classical fluid affect the large scale wind?
\end{enumerate}
We will argue that the answers to both the above the questions may indeed be negative, motivating a set of simulations of a highly simplified ISM. Such a simulation would also improve our physical understanding of the role of the individual processes.

On the first question we note that the progenitor of type II core collapse SNe are massive stars \citep{Smartt_2009} with lifetimes $\sim 1-30$ Myrs  \citep{Portinari_1998}, therefore the majority of SN energy associated with an instantaneous burst of stars with for example a \cite{Chabrier_2003} initial mass function will be released within $\sim 30$~Myrs. It is thought the birth cloud of such stars is likely destroyed before by the combined effects of stellar winds, proto-stellar jets and radiation (e.g. \citealp{Matzner_2002}), and there is observational evidence for this (e.g. \citealp{Lopez_2011}). Some clouds may form by turbulent compression when overrun by a spiral arm, and may disperse by the same flows that created them in the first place on a short time-scale (\citealp{Dobbs_2008}, see also \citealp{Tasker_2009}). 

In any case, when the SNe explodes it will in general not do so inside its natal cloud. For this reason we assume that the SNe explode in typical environments in the disk plane of galaxies. Note however that the SNe may still be clustered rather than Poisson, a complication that we neglect. Typical giant molecular clouds have a velocity dispersion of $\sim 4$ km $\rm s^{-1}$ \citep{Scoville_1987, McCray_1987}, which over 10 Myr results in a dispersion of around $40$ pc, which is a significant fraction of our box size and the typical distance between molecular clouds.

The second question is delicate, and worthy of significant discussion. We first note that we follow the nomenclature of \citet{Wolfire_1995}, where the $T\sim 100$K phase of the ISM is called the cold neutral medium (hereafter CNM), the $T\sim 10^4$K phase as the warm neutral medium (hereafter WNM) and the $T \gtrsim 10^6$K phase as the hot ionised medium (HIM). The CNM exists in the form of dense clouds, occupying a very small fraction of the total volume but with a significant fraction of the total mass. These clouds are believed to be in pressure equilibrium \cite{Spitzer_1956}, with the WNM and HIM, thus making their energy budget (pressure $\times$ volume) also a small fraction of the ISM thermal energy. Their pressure support is probably composed of a combination of magnetic, thermal and cosmic ray components. The proportions of energy in thermal, bulk and turbulent motions of the HIM and WNM are still not entirely known though there is consensus that much of the turbulence is supersonic \citep{Elmegreen_2004}. A supersonic nature of turbulence in the ISM requires that the energy budget is dominated by inertial terms of the turbulent motions over the thermal and magnetic terms in the WNM and HIM.

Despite their small fraction of the energy budget, however, the cold phase can perform the role of a heat sink. Thermal energy from the warm and hot phases can be transported in to the cold phase via thermal conduction which can be dissipated via the molecular transitions of this cold gas (particularly CO, ${\rm H}_2$), metal lines (in particular CI*), and dust.  The excited states of the molecules, however, are rather long lived and whilst they are certainly important for star formation they may not significantly cool the WNM phase of the ISM due to its sparse nature \citep{deJong_1980,Martin_1996}. The molecules also play an important role as an absorber of photo-ionizing radiation, however we will ignore radiative driving here. The simulations described in this paper simply neglect the cold phase, by truncating the cooling function below a value of $T_0 = 10^4$~K. If we were to include cooling below 
$T_0$ we would have to include significantly more physics (magnetic fields, heat conduction, diffuse heating): here we want to investigate and understand the simpler yet still complex case of a two-phase medium.

We have also intentionally left out the physics of cosmic rays (see, e.g. \citealp{Pfrommer_2007}) and magnetic fields (e.g. \citealp{Breitschwerdt_2007}) which may be important in providing support against collapse, particularly in the CNM. Our goal is to understand the resultant ISM without these complications, before discussing the implications of their addition. We would also like to stress that although we have included gravity, we have not included \emph{self}-gravity (i.e our gravity is time-independent and only self-consistent for the initial set-up) which would be a poor approximation if we had included the dense, cold material of the CNM. Without the CNM gravity does not influence material on scales below the Jeans length of the WNM, equivalent to the scale height of the warm disk. 

\section{Simulation set-up}
\label{sect:simulations}
In the following section we will describe the simulations we have performed of supernova driven outflows from an idealised ISM. Our simulations model 
the ISM and halo of a disk galaxy in a tall column, with long ($z$) axis perpendicular to the galactic disk, and co-rotating with the disk material. We use outflow conditions at the top and bottom of the column, and periodic boundary conditions in $x$ and $y$. We describe the initial conditions of the gas and the physical processes (gravity, cooling and supernova feedback) we have included, and detail their numerical implementation. Finally we describe some tests we have performed on the code and the parameters we chose to explore in our simulation set.

The simulations use a modified version of the  \flash\ 3 code  \citep{Fryxell00}. \flash\ 3 is a parallel, block structured, uniform time-step, adaptive mesh refinement (AMR) code. Its second order (in space and time) scheme uses a piecewise-parabolic reconstruction in cells. Due to the extremely turbulent nature of the ISM in our simulations, we find that \flash\ attempts to refine (i.e. to use the highest resolution allowed) almost everywhere within our simulation volume. Therefore we disable the AMR capability of \flash\ and run it at a constant refinement level (albeit varied for our resolution studies). To mitigate the overhead of the guard-cell calculations we increase our block size to $32^3$ cells per block.

For the gas physics we have assumed a monatomic ideal gas equation of state,
\begin{equation}
p = (\gamma - 1) \rho u\,,
\end{equation}
where $u$ is the specific thermal energy and $\gamma=5/3$ is the adiabatic index. This deviates slightly from the physical equation of state which should include the transition in mean particle mass that occurs as the atomic hydrogen becomes ionised, but the impact of this simplification is small compared to the other uncertainties in this kind of simulation.

\subsection{Physical processes}\label{sec:PhysProc}
The simplified ISM discussed in Section \ref{sec:SelfConsist} is shaped by three fundamental processes: gravity, cooling and energy injection from supernovae, which dominate when we are only considering the WNM and HIM. We stress that our aim is to simplify the problem as much as possible so that we can extract the physical principles. In future works we will experiment with making some assumptions more realistic. Below we discuss the effects and implementation of all these processes.

\subsubsection{Gravity}
The gas in our simulations is initially in (vertical) hydrostatic equilibrium. In a disk galaxy the gravitational acceleration is induced by the gas and stars in the disk, baryons in the bulge and dark matter (in the halo and possibly the disk, see e.g. \citealt{Read_2008}). Despite these complications, when one moves to the (non-inertial) frame moving locally with the disk, the dominant effective potential lies in the vertical direction, with a scale height of a few hundreds of parsecs. Since the shape of this profile is approximately in accordance with the gaseous one, we model the total gravity of all components (gas, stars, dark matter) as being in proportion to the gaseous component, with a multiplier of the inverse of the gas fraction, $\frac{1}{f_{\rm g}}$, to account for the stellar and dark matter components, i.e. the gravitational potential depends on the gas density through Poisson's equation as
\begin{equation}
f_{\rm g} \nabla^2 \phi = 4 \pi G \rho \label{eq:poisson} \; .
\end{equation}

We also make a second assumption, namely that the gravitational profile of the disk is fixed in time, $\phi = \phi [ \rho_0 ]$. This is assumed because the minimum temperature of our cooling function (discussed in Section \ref{sec:cooling} below) sets the Jeans length on the order of the scale of the disk height, so we do not expect smaller self-gravitating clouds to appear in our simulations. In contrast, in the ISM of the Milky Way small self-gravitating clouds can form, because the ISM does cool to lower temperatures, however the physics of star formation is not the process we wish to address in this paper.

Other terms we have neglected include those introduced by the Coriolis force across our simulation volume, due to our non-inertial choice of frame,
\begin{equation}
\dot{\bf v}_{\rm cor} = -2{\bf \Omega} \wedge {\bf v}\, ,
\end{equation}
where $\Omega$ is the angular velocity of the galaxy. Our simulations, however, will typically be of such short time scales and volumes that the Rossby number (the ratio of inertial to coriolis terms) is large. Nevertheless, more complete simulations would include this, along with the time dependent gravitational changes introduced by spiral density waves. Note that our simulations also neglect the velocity shear that is present in a differentially rotating disc.

\subsubsection{Radiative cooling}\label{sec:cooling}
The cooling function $\Lambda(T)$ of $T\sim 10^4-10^7$K gas with solar abundances is primarily due to bound-bound and bound-free transitions of ions, whereas above $T=10^7$K it is largely dominated by bremsstrahlung \citep{Sutherland_1993}. Below $T\sim 8000$K there is a sharp decrease by several orders of magnitude, causing a build up of gas in the WNM. Cooling below $8000$K is due to dust, metal transition lines such as CI*, and at very low temperatures, molecules. 

Whilst the imprint of small features in the cooling function should be observable in the ISM, it is really the cut-off at $T\sim 8000$~K that controls the WNM, and as such we approximate the cooling function with a Heaviside function with a step at $T_0 = 10^4 \; \rm K$,
\begin{equation}
\rho \dot{u} = \left\{ \begin{array}{cc} 
 -\Lambda n^2 ,&  T \geq T_0 \\
0 , & T< T_0\,,  \end{array} \right.
\label{eq:cf}
\end{equation}
where we in addition assume pure hydrogen gas so that the number density $n = \rho / m_p$, and $\Lambda=10^{-22} \, \rm erg \, cm^3 \, s^{-1}$ (although it is varied in a few of the simulations). We implement this very simple functional form so that we can explicitly check the effect of the normalisation of the cooling function, and to make sure that any characteristic temperature of the gas is not due to features in $\Lambda$.

The cooling function of Eq.~(\ref{eq:cf}) results in a cooling time for gas with $T \geq T_0$ of
\begin{eqnarray}
t_{\rm cool} &\equiv & \frac{m_p  u}{ \Lambda n} \nonumber\\
&\approx& 660 \, {\rm yr} \left( \frac{T}{T_0} \right) \left( \frac{n}{1 \; {\rm cm}^{-3}} \right)^{-1} \times \nonumber\\
& &  \left( \frac{\Lambda}{10^{-22} \, \rm erg \, cm^3 \, s^{-1}} \right)^{-1} \; .
\end{eqnarray}
Since we have chosen a discontinuous function for our cooling, we implement a scheme in our code which prevents cooling below $T_0$ (although the hydrodynamic forces can still achieve lower temperatures adiabatically). This largely prevents the overshoot errors resulting from an explicit solver in this kind of problem.

To test the importance of the choice of cooling function, we also implemented the cooling function appropriate for cosmic gas with solar abundance pattern from \citet{Sutherland_1993},
\begin{equation}\label{eq:SD_cooling}
\Lambda_{\rm SD} (T) = 5.3\times 10^{-24} \left( T_8^{1/2} + 0.5 f_m T_8^{-1/2}  \right) \, \rm erg \, cm^3 \, s^{-1} \, ,
\end{equation}
where $T_8\equiv T/10^8~{\rm K}$, with $f_m=0.03$ for low metallicity gas, and $\Lambda =0$ for $T<10^4\, \rm K$. All runs where this cooling function  is used are marked `SD' (see table \ref{tab:parameters}). The minimum of this cooling function is at $5\times 10^7 f_m\; \rm K$, where the cooling rate 
\begin{equation}\label{eq:SD_min}
\Lambda_{\rm SD,min} = 1.30 \times 10^{-24} \, \rm erg \, cm^3 \, s^{-1} \, ,
\end{equation}
(ignoring the cut-off below $10^4\; \rm K$). We show in Appendix \ref{sec:convergence} that the behaviour of the ISM in our simulations is surprisingly independent on the exact shape of the cooling function $\Lambda(T)$, although it depends on the minimum value at high temperatures $\gg 10^4$~K.

\subsubsection{Energy injection by supernovae}
The Kennicutt-Schmidt (KS) relation connects observed surface density of star formation in a disk galaxy, $\dot{\Sigma}_\star$, to its gas surface density
$\Sigma_{\rm g}$,
\begin{equation}
\dot{\Sigma}_\star \approx  2.5\times 10^{-4} \Sigma_{\rm g1}^{1.4} \, \rm M_\odot \, yr^{-1} \, kpc^{-2} \,,
\label{eq:KS}
\end{equation}
\citep{Kennicutt_1998}, where $\Sigma_{\rm g1} \equiv\Sigma_{\rm g} / 1 \, \rm M_\odot pc^{-2}$. We also perform some simulations with an alternative formulation using a higher star formation rate, more commonly used in cosmological simulations, discussed in Appendix \ref{sec:convergence}. Notably this introduces an additional dependence on the gas fraction of the disk, $f_{\rm g}$, that is absent from the KS relation. Our idealised model of a supernova explosion is the injection of  $10^{51}$ ergs \citep{Cox_72} of thermal energy in a small volume, implicitly assuming instantaneous thermalisation of the SN ejecta.  The distribution in time of these is taken to be a Poisson process (the Poisson process has the Markov property and so our SNe are independent) with a time independent rate computed from the initial parameters of the disk.  For the local spatial distribution of SNe we assume the star formation rate to be proportional to the initial density, i.e.
\begin{equation}
\mathbb{E} [ \dot{\rho}_\star {\rm \, dV \, dt} ] = \dot{\Sigma}_\star \frac{\rho(t=0) }{\Sigma_{\rm g}} {\rm \, dV \, dt}\, .
\end{equation} 
A consequence of this choice is that if the scale height of the gas profile evolves significantly the distribution of SNe will become inconsistent with the instantaneous mass profile. We discuss this further later.

Given the star formation rate, the associated core-collapse SN rate is computed assuming the stellar initial mass function yields
$\varepsilon_{100}$ SNe per $100 \, \rm M_\odot$ of star formation. For reference, for a \cite{Chabrier_2003} initial mass function with stars with masses $\in[0.1,100]\; \rm M_\odot$, of which those with mass in the range $[6,100]\; \rm M_\odot$ undergo core collapse, $\varepsilon_{100}=1.8$.

The final element of the SN prescription is the distribution of the injected energy over the computational grid. The choice of volume over which to spread the thermal energy of the supernovae is influenced by two considerations. If the volume is too large the remnant will evade the adiabatic expansion phase and immediately proceed to the radiative phase \citep{Cox_72, Creasey_2011}. If the volume is very small the code will require many extra time steps evaluating the initial stages of a Sedov-Taylor blast wave and will perform unnecessary computation\footnote{To get some idea of the computational requirement of this, we recall that the velocity of a 3 dimensional Sedov blast wave evolves as $v \sim t^{-3/5}$. Substituting this into the Courant-Friedrichs-Lewy (CFL) condition we see that the number of time steps required to reach a given radius is proportional to that radius.}. Following \citet{Cox_72}, the radius at which the blast wave cools and forms a dense shell is 
\begin{eqnarray}
R_s &=& 15.6 \left( \frac{E_{\rm SN}}{10^{51} \, \rm erg} \right)^{3/11} \left( \frac{\Lambda}{10^{-22} \, \rm erg \, cm^{3} \, s^{-1}} \right)^{-2/11}  \times \nonumber \\
&& \left( \frac{n}{1 \, \rm cm^{-3}} \right)^{-5/11} \; \rm pc \, , \label{eq:r_shell}
\end{eqnarray}
however to account for higher densities and the numerical spreading of shocks it is wise to resolve a fraction of this \citep{Creasey_2011}. 

Taking the above into consideration, for our simulations we spread the thermal energy of each SN over several cells given by the multivariate (3D) normal distribution of standard deviation 2 pc, consistent with being smaller than the cooling radius of Eq.~(\ref{eq:r_shell}) for densities $n < 77 \, \rm cm^{-3}$ ($\rho< 1.3\times 10^{-22}$~g~cm$^{-3}$).

\subsubsection{Time-stepping}

In addition to the numerical considerations described above, we also needed to make some adjustments to the time step calculation in \flash. The default time-stepping scheme in \flash\ uses a Strang-split method (\citealp{Strang_1968}, an operator splitting method where the hydrodynamic update occurs in two half steps, with the order in which the Riemann-solver operates reversed from $xyz$ to $zyx$ between the first and the second half step). Source terms such as the injection of SN energy, are evaluated at the end of each half step, after the Riemann solver has been applied. This makes the implementation of the supernova energy injection problematic, as the thermal energy in a cell can increase by many orders magnitude followed by a hydrodynamic step before a new time step is calculated. The latter hydrodynamic step then almost inevitably violates the CFL condition and the Riemann solver fails to converge. We avoid this by making the timestep limiter for the supernova source terms {\it predictive}, i.e. we utilise the foreknowledge of the pre-computed SNe times to recognise when a supernova will occur before the end of the timestep given by the CFL condition and return a timestep of either up to just before the supernova, or of the predicted CFL timestep after the supernova has occurred, whichever is smaller.

It is worth contrasting this with some other simulations of the ISM. In a series of simulation \citet{Avillez_2004, Avillez_2005a, Avillez_2005b} uses a set up similar to ours, with imposed gravity, cooling, SNe turbulence and magnetic fields in columns through disks of $1\times 1 \times 10$ kpc, although the focus is not on the mass loading. More recently the ERIS simulations \citep{Powell_2011} simulated the ISM in a single high redshift dwarf galaxy.  \citet{Cooper_2008} perform a simulation of the central region of an M82-like starburst galaxy with gravity, cooling and energy injection due to supernovae (although this energy injection is continuous within a central volume, rather than stochastic as in our simulations).

\subsubsection{Code tests}
A set up as complex as this requires some testing to confirm that the physical processes have been correctly implemented. As such we ran a number of simpler problems as well as the convergence tests in Appendix \ref{sec:convergence}.

In order to test our hydrostatic set up we simulated the disk without supernovae for several dynamical times. Some sub-percent evolution in the gas occurred, almost certainly due to our evaluation of the analytic solution for the gravitational potential and density at the centres of cells producing some discretisation error. The implementation of the cooling function was tested largely in \citet{Creasey_2011}. We follow a similar approach where we made the cooling rate for each cell an output of our code which was compared with the instantaneous rate predicted from the temperature and density of each cell (again there were small differences due to the comparison of an instantaneous rate with the average from an implicit scheme).

The implementation of the individual SN in our set-up is largely similar to that of the Sedov-Taylor blast wave solution implemented in \flash\ as a standard test, and we compared it to the similarity solution. We calculate the location and times of SNe explosions ahead of the simulation, and verify that the code indeed injects them correctly.

We initially also performed these calculations using the \gadget\  simulation code \citep{Springel_05} that has been successfully applied to many cosmological simulations. Unfortunately the adaptive time-stepping algorithm proved problematic for correctly following the blast waves, and we noticed similar problems as recently highlighted by \cite{Durier_Dalla_Vecchia_12}: particles may be on long time-steps in the cold ISM, and largely fail to properly account for being shocked by the blast wave from a nearby particle. \cite{Durier_Dalla_Vecchia_12} addressed this problem with a time step propagation algorithm, however we did not have this nor the algorithm of \cite{Saitoh_2009} available and the alternative of a global timestep would have been far too computationally expensive due to the large dynamic range in time steps required in the evolution of the blasts. As such we used the global adaptive time stepping algorithm of \flash . 

\subsection{Initial conditions}
Our initial setup is a tall box poking vertically through an idealised disk profile. We choose the long axis in the $z$-direction in order to capture a multiple of the gravitational scale height of the disk. The profile is a 1-dimensional gravitationally bound isothermal one with gas surface density $\Sigma_{\rm g}$. As discussed in section \ref{sec:PhysProc} we have excluded the effects of shear (due to the Coriolis force in the disk) and large scale motions which may drive some turbulence down to the small scales. The gas density is 
\begin{equation}
\rho(z) = \frac{\Sigma_{\rm g}}{2b} {\rm sech}^2 \left(\frac{z}{b}\right)\,,
\end{equation}
and the corresponding gravitational acceleration follows from Eq.~(\ref{eq:poisson}), 
\begin{equation}
\nabla \Phi = 2 \pi G \Sigma_{\rm g} f_{\rm g}^{-1} {\rm tanh}  \left(\frac{z}{b}\right) \; .
\end{equation}
Setting the gas temperature to $T_0$ (which is also the base of the imposed cooling function) and assuming the gas to be initially in hydrostatic equilibrium, the scale height is
\begin{eqnarray}
 b &=& \frac{ f_{\rm g} k_{\rm B} T_0}{{\rm m_p}\pi G \Sigma_{\rm g}} \\
&\approx& 61  \left( \frac{ f_{\rm g}}{0.1} \right) \left( \frac{ \Sigma_{\rm g}}{10 \, \rm M_\odot \, pc^{-2}} \right)^{-1} \; {\rm pc}\,, \label{eq:b}\\
\end{eqnarray}
where numerically
\begin{equation}\label{eq:density}
\rho(z) \approx 3.4 \, \left( \frac{ \Sigma_{\rm g}}{10 \, \rm M_\odot \, pc^{-2}} \right)^2 \left( \frac{ f_{\rm g}}{0.1} \right)^{-1} {\rm sech}^2 \left(\frac{z}{b}\right)  \, {\rm m_p \, cm^{-3}}\,.
\end{equation}

The (vertical) dynamical time of the disk is
\begin{eqnarray}
t_{\rm dyn} &=& \sqrt{\frac{b f_{\rm g}}{G \Sigma_{\rm g}}}\nonumber \\
& \approx & 12 \times 10^6  \left( \frac{ f_{\rm g}}{0.1} \right) \left( \frac{ \Sigma_{\rm g}}{10 \, \rm M_\odot \, pc^{-2}} \right)^{-1} \, {\rm yr}\,,
\label{eq:tdyn}
\end{eqnarray}
and the ratio of the dynamical time to the cooling time
\begin{eqnarray}
\zeta &\equiv& \frac{t_{\rm dyn}}{t_{\rm cool}} \nonumber \\
&\approx& 1.7 \times  10^5 \left( \frac{ \Lambda}{10^{-22} \, \rm erg \, cm^3 \, s^{-1}} \right) \left( \frac{ \Sigma_{\rm g}}{10 \, \rm M_\odot \, pc^{-2}} \right) \, .
\end{eqnarray}

The exact gravitational potential is given by
\begin{equation}\label{eq:gravpot}
\Phi(z) = 2 \pi G b \Sigma_{\rm g} f_{\rm g}^{-1} \log {\rm cosh}  \left(\frac{z}{b}\right) \, ,
\end{equation}
and the pressure in hydrostatic equilibrium is
\begin{eqnarray}
p &=& \pi G \Sigma_{\rm g} f_{\rm g}^{-1} b \rho(z) \\
&\approx & 3.3 \times 10^4  \left( \frac{ \Sigma_{\rm g}}{10 \, \rm M_\odot \, pc^{-2}} \right)^{2} \times \\
&& \left( \frac{ f_{\rm g}}{0.1} \right)^{-1} {\rm sech}^2 \left( \frac{z}{b} \right) \, \rm K \, cm^{-3} \, .
\end{eqnarray} 
Finally, the hydrostatic temperature for all our disks is chosen to be
\begin{equation}
T_0 = 10^4 \, \rm K \, .
\end{equation}

\subsection{Numerical parameters and boundary conditions}
\begin{table}
\begin{center}
\begin{tabular}{ | r | c | c |}
 & & Fiducial \\
 & Range of values explored & value \\
\hline
$\Sigma_{\rm g}\, ({\rm M_\odot \, pc^{-2}})$ \vline & 2.5, 3.23, 4.17, 5.39, 6.96, & 11.61   \\
\vline &  8.99, 11.61, 15, 30, 50, 150, 500 &  \\
$f_{\rm g}$ \vline & 0.01, 0.015, 0.022, 0.033, & \\
\vline & 0.050, 0.1, 0.2, 0.5, 1.0  & 0.1 \\
$\dot{\Sigma}_\star$ \vline & Eq. (\ref{eq:KS}), (\ref{eq:KStdyn}) & Eq. (\ref{eq:KS}) \\
$\Lambda\, ({\rm erg \, cm^3 s^{-1}})$ \vline & 1, 2, 4, 8, 16$\times10^{-22},\, \rm SD$  & $10^{-22}$ \\
Resolution (pc) \vline & 0.78, 1.56, 3.12, 6.25 & - \\
\end{tabular}
\caption{Parameter variations in our simulation. Each simulation is initialised with an isothermal profile with a surface density of $\Sigma_{\rm g}$ in cold gas and gas fraction of $f_{\rm g}$ (i.e. a total mass density of $\Sigma=\Sigma_{\rm g}/f_{\rm g}$). Star formation proceeds either in a pure Kennicutt-Schmidt prescription, or the dynamical time variation in Eq.(\ref{eq:KStdyn}). Cooling above $10^4\; \rm K$ proceeds at a rate $\Lambda$ and we study the simulations at several resolutions to test for convergence.}
\label{tab:parameters}
\end{center}
\end{table}

To produce simulations of a realistic ISM we make the following choices of parameters. In terms of resolution we must have cell sizes fine enough to capture the cooling of supernova remnants (Eq.~\ref{eq:r_shell}) yet the simulation volume needs to be large enough to capture several scale heights of the star forming disk. In terms of gas fractions and gas surface densities we choose values approximating those in the solar neighbourhood and some variations. In practise we chose fiducial values for the disk parameters ($\Sigma_{\rm g} = 11.61 \, \rm M_\odot\,  pc^{-2}$, $f_{\rm g}=0.1$) and examine this reference model in detail. For reference, the gas surface density of the solar neighbourhood of the Milky Way has been estimated at $\Sigma_{\rm g} = 13.2 \rm \; M_\odot \, pc^{-2}$, with a dynamical density of $\Sigma_\star = 74 \; \rm M_\odot \, pc^{-2}$ \citep{Flynn_2006}.

In order to test the dependence of winds on the disk properties we perform a slice of the parameter space varying $\Sigma_{\rm g}$ and $f_{\rm g}$ (see Table \ref{tab:parameters}). Not all parameter combinations are explored, as we cut out the simulations with very small scale heights (due to resolution constraints) and large scale heights (due to the finite box size). The dependence of the results on cooling, resolution, box size, star formation rate and run time can be seen in the Appendix.

All our simulations were conducted in box sizes of $200\times 200 \times 1000 \; \rm pc$ with constant cell sizes. All cells were cubic, and in the vertical direction the number of cells for our default resolution is 640, with corresponding cell size of 1.6~pc. We vary the numerical resolution
using 160,320,640,1280 cells in $z$, with corresponding cell sizes ranging from $6.25-0.78$ pc. These simulations are denoted L2, L3, L4, L5 respectively. We also test the effect of adjusting our box size with simulations of $2 \times$ and $4 \times$ the width (see Appendix \ref{sec:convergence}).

The gas surface density $\Sigma_{\rm g}$ is varied from $2.5 M_\odot \, \rm pc^{-2}$ to $15 M_\odot \, \rm pc^{-2}$ in $8$ logarithmically spaced steps followed by three additional steps of 30, 50 and 500 $\rm M_\odot\, pc^{-2}$ . Notably some of these are below the minimum surface density threshold for star formation of \citet{Schaye_2004} of $3-10\; \rm M_\odot \, pc^{-2}$ (although there is evidence that star formation proceeds below this level, e.g. \citealp{Bigiel_2008}). The gas fraction $f_{\rm g}$ was varied from $0.01$-$0.05$ in 5 logarithmic steps followed by additional steps of $0.1$, $0.2$, $0.5$ and $1.0$. The cooling function $\Lambda$ was varied from $3.9\times 10^{-25}$ to $1.6\times 10^{-21}\; \rm erg \, cm^3\, s^{-2}$, and we ran additional models with the \citet{Sutherland_1993} cooling function as parameterised in  Eq. (\ref{eq:SD_cooling}). Each of our experiments is evolved over 20~Myr (typically thousands of cooling times) in order to simulate many SNe.

\section{Results}
\label{sect:results}
In this section we discuss the results of the simulations described in the previous section. We begin with a discussion of a single snapshot, allowing us to investigate the instantaneous properties of the idealised ISM and outflow. We then move to looking at the evolution of a simulation and the statistics we can measure before finally investigating the effects of all the parameters discussed in the previous section.

\subsection{Fiducial run}

The impact of SNe depends strongly on whether they explode in the dense gas or in the more rarefied HIM. The supernovae in the disk blast bubbles in the ISM and compress the warm gas into thin sheets and clouds. We note that between the different simulations the volume of the warm medium can vary from a series of disconnected, nearly spherical regions to a highly porous stratus that approximately covering the base of the disk potential. We will use the term `clouds' to apply to both. When supernovae explode in the rarefied regions, either at the edge of the disk or inside previously evacuated bubbles, the heated gas pushes out of the central region and then rapidly escapes from the simulation volume in a zone of acceleration above and below the disk. This is the ISM portion of the galactic wind (i.e. the gas whose thermal energy far exceeds the potential barrier to escaping the disk). Some warm clouds are dragged along with this wind. A movie of this simulation is available online along with time dependent versions of some of the other figures
\footnote{See \url{http://astro.dur.ac.uk/~rmdq85}}

\begin{figure*}
\centering
\includegraphics[width=2\columnwidth]{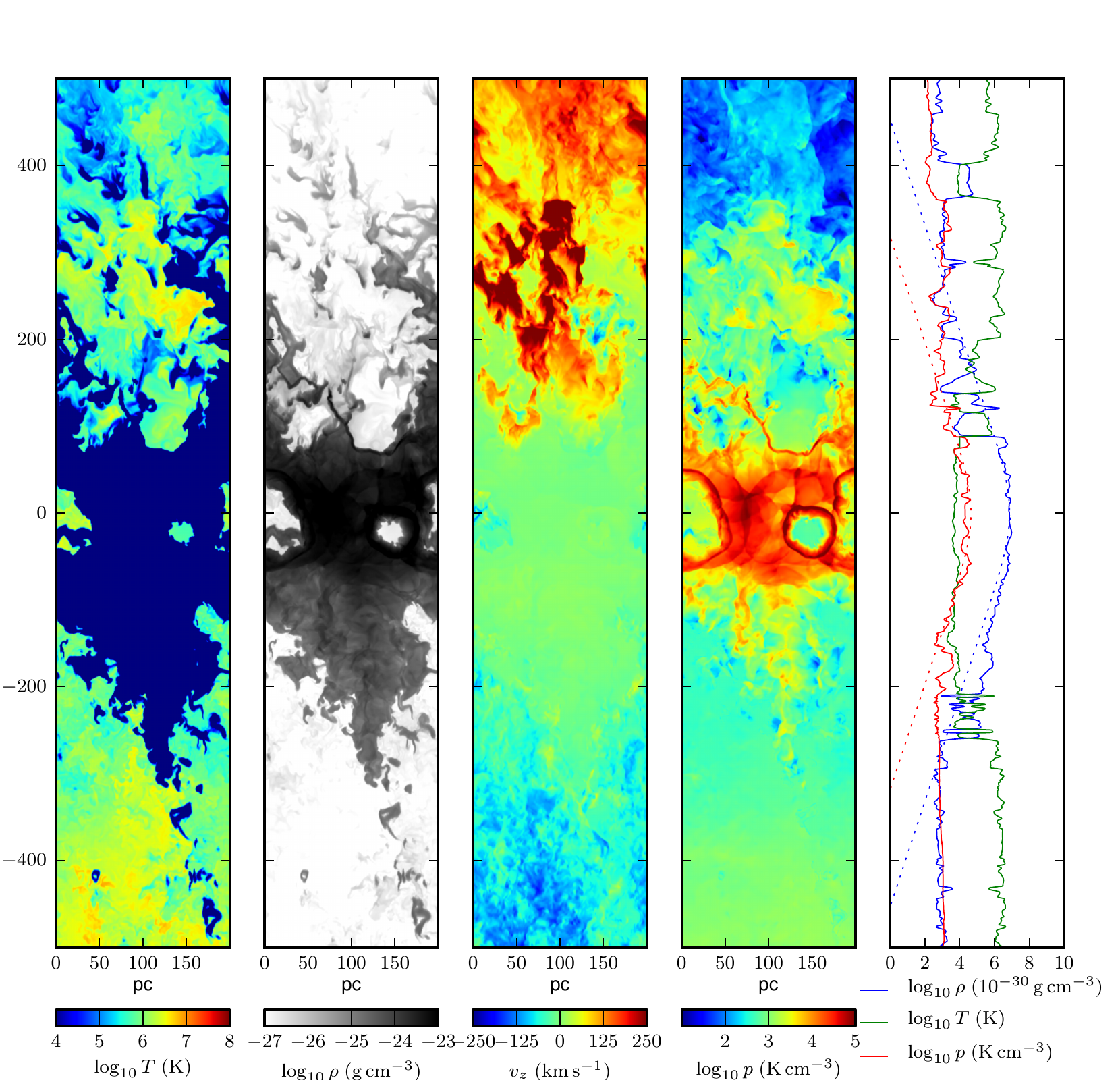}
\caption{Left to right, temperature, density, vertical velocity and pressure plots through a slice of the simulation, at time 5 Myr. Temperature is coloured between $10^4-10^8 \, \rm K$, density between $20^{-27}-10^{-23} \; \rm g\,cm^{-3}$, $v_z$ from $-250$ to $250\, \rm km \, s^{-1}$ and pressure from $10-10^5\;\rm K\,cm^{-3}$. On the far right is the profile of density, temperature and pressure along a vertical line through the centre of the slice. In \emph{dotted blue} and \emph{red} we show the hydrostatic density and pressure profiles at $t=0$. Around $z=0$ we can see the disrupted disk in the temperature and density plots, with the warm gas squeezed into sheets and globules, and a significant fraction of the volume now consumed by a hot ($\sim 10^{6.5} \, \rm K$) sparse phase. In the velocity plot we can see a bulk vertical outflow from the disk. The outflow is inhomogeneous, entraining significant turbulence as well as some warm gas, swept away from the disk. }
\label{fig:temp_dens_vel}
\end{figure*}

\begin{figure}
\centering
\includegraphics[width=\columnwidth]{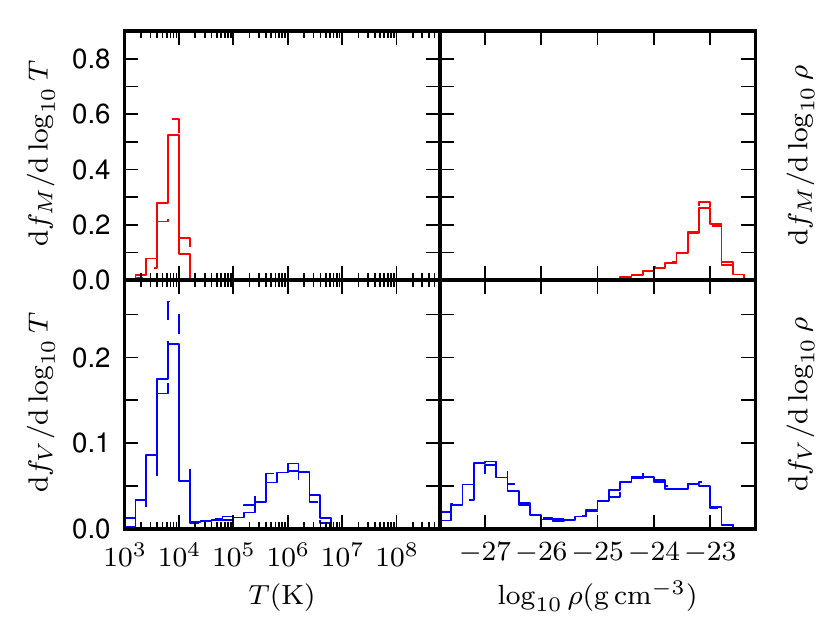}
\caption{Density and temperature probability distributions for the fiducial run at 10 Myr, as shown in Fig.~\ref{fig:temp_dens_vel}, \emph{solid, dashed} and \emph{dotted} lines denote the L4, L3, L2 resolution runs, respectively. \emph{Upper panels} show the mass fractions in temperature and density, \emph{lower panels} show the corresponding volume fractions. We see a clear bimodality between the WNM (at low temperature and high density) and the HIM (at high temperature and low density). Almost all of the mass is in the WNM phase, but a significant fraction of the volume in the HIM.}
\label{fig:volfrac}
\end{figure}

In Figure \ref{fig:temp_dens_vel} we show an $x-z$ slice of the fiducial run, at a time of 12 Myr. We can see that the combined action of multiple SNe has disrupted the disk considerably, with the warm gas squeezed into dense sheets and globules entrained in outflowing gas, and around half the volume now occupied by a hot tenuous phase. The gas appears to be in well defined phases, an HIM (greens and yellows) and a WNM (dark blue) with little gas at intermediate temperatures (sell also Fig. \ref{fig:volfrac}). Notably there is more temperature variation in the hot phase (a few orders of magnitude) than in the WNM (which is all close to $10^4 \; \rm K$). The density plot also appears to show two distinct phases, a high and low density medium, where the high densities show up in the temperature plots as WNM. In the velocity plot we can see a bulk vertical outflow from the disk, with velocity correlating with height. The pressure plot shows a dramatically lower dynamic range than either the temperature or density plots, but has some distinctive shells due to individual SN remnants. The impression of a volume in quasi pressure equilibrium is reinforced by the profile plot where the temperature and density fluctuations appear to anti-correlate, resulting in comparatively small pressure variations.

Above the plane of the disk the outflow is also very inhomogeneous, containing significant turbulence as well as some warm clouds or globules with cometary shapes. The corresponding locations in the density and pressure panels reveal that these clouds are also overdense and slightly under-pressured. In velocity the clouds appear to be receding from the disk at a lower velocity than the HIM, that is rushing past them at around 100~km~s$^{-1}$. The hot wind is stripping the edges of these warm clouds, as evidenced by their tails (see also the movie online).

After only 12 Myr the original disk has undergone considerable disruption but is still observed as a connected feature in this slice (and the majority of the mass of the simulation remains in the central region). The disk has also been disrupted asymmetrically, with more mass pushed into the lower half space by the stochastic locations of the SNe. The externally imposed gravity will ultimately return this mass to the base of the potential, yet the combined action of the supernovae has been enough to displace it.

Whilst we have run these simulations at different resolutions, it is important to note that the turbulent and chaotic nature of these simulations results in specific features such as individual clouds being at different locations or indeed absent between the different runs. Global properties, however, such as the outflow mass and temperature will be less stochastic, and we devote Appendix \ref{sec:convergence} to the convergence study of these properties. 
In general these simulations are numerically well converged. In the following figures we also include a few convergence comparisons where space allows.

The value of the ISM pressures in our simulations are around $10^3 \, \rm K \, cm^{-3}$, comparable to the pressure in simulations such as \citet{Joung_MacLow_2006} and \cite{Joung_MacLow_2009}. Estimates of the pressure of a star forming ISM vary, \citet{Bowyer_1995} find a pressure of around $2\times 10^4\, \rm K \, cm^{-3}$ in the local bubble, although in the centre of the highly star forming region of 30 Doradus, \citet{Lopez_2011} estimate a pressure of $\sim 7 \times 10^6 \, \rm K \, cm^{-3}$ from IR dust measurements.

Figure \ref{fig:temp_dens_vel} suggests that the hot and warm phases are quite distinct, and we test this by inspecting the volume fractions in Fig.~\ref{fig:volfrac}. The warm phase is very tightly distributed below $10^4$K, as we might expect since the only mechanism for cooling here is by adiabatic expansion. The lack of intermediate temperatures suggests they have very short cooling times, which is consistent with a pressure equilibrium view. The hot tail of the distribution suggests the hottest gas either mixes with cooler gas or escapes from the simulation volume.

\begin{figure}
\centering
\includegraphics[width=\columnwidth]{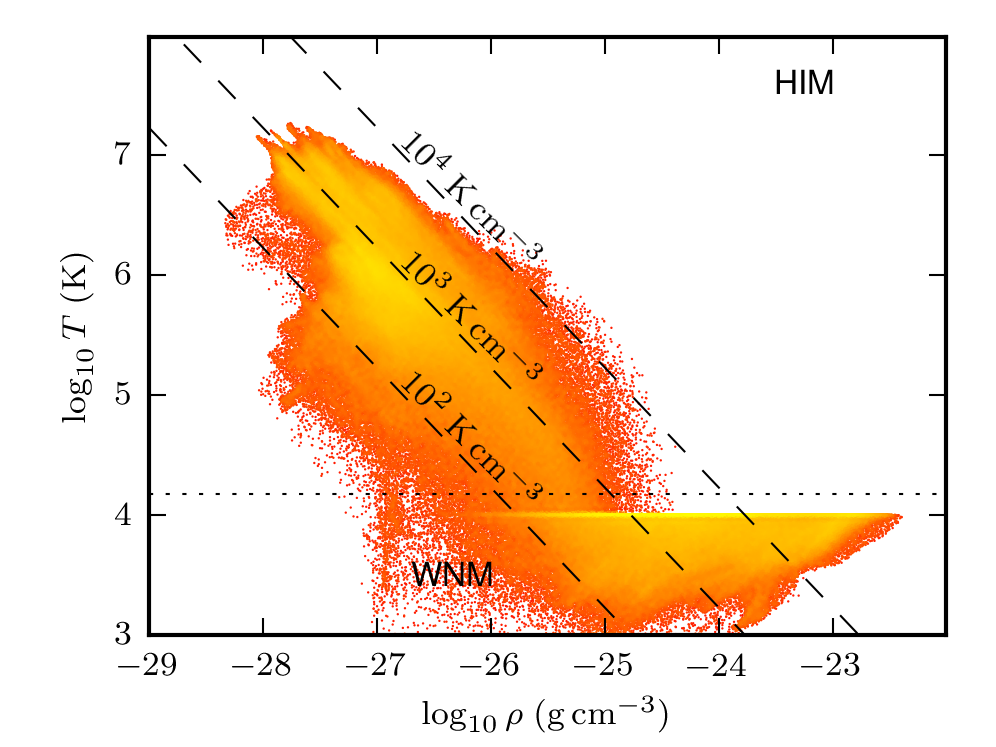}
\caption{Density-temperature histogram for the fiducial model at L3 resolution. Each pixel is coloured by the fraction of cells at given $\rho-T$. \emph{Dashed black lines} indicate lines of constant pressure, $p/k_{\rm B}=10^2,10^3,10^4 \; \rm K \, cm^{-3}$ as indicated in the panel. We see the simulation volume is in an order-of-magnitude pressure equilibrium, with a bimodality in the gas phases into an HIM and WNM that we have segregated approximately with the \emph{dotted black line},  a temperature cut at $15,000 \, \rm K$. Above $10^4$K and $\rho > 10^{-24} \rm g\, cm^{-3}$ the cooling time of the gas is very short and the gas quickly cools to $10^4$K. Some gas reaches lower than this temperature due to adiabatic expansion.}

\label{fig:phase}
\end{figure}

Figure \ref{fig:phase} is the density-temperature phase diagram for the fiducial model at L3 resolution (3~pc cells), which is broadly described by two regions. In the lower right, lying horizontally at a nearly constant temperature of order $T_0=10^4$~K (the base of the cooling curve) is the WNM, which contains most of the mass. The HIM is in the upper left. On examination of time dependent movie of this simulation we see the structure in the HIM is due to multiple supernovae, each supernova blast forms a `finger' roughly along an isobar, and as these shocked regions evolve and expand these lines descend to lower temperatures forming the mixture in the lower right region. As one looks to lower temperatures the fingers start to merge and become indistinct. We see that instantaneously we have variations in pressure within approximately one order of magnitude, and that a significant fraction of the volume is in the HIM.

\subsubsection{The characteristic temperature of the HIM}\label{sec:CharacTemp}
It is interesting to consider where the characteristic temperature of the hot phase may appear from. We recall that the cooling function used in these simulations was intentionally chosen to be independent of temperature for $T\ge T_0=10^4$~K, and as such cannot by itself introduce a characteristic temperature scale, yet in Fig.~\ref{fig:volfrac} the hot gas quite clearly has a well defined peak temperature $\sim 10^{6}\; \rm K$. This is much higher than the escape temperature for the simulation volume ($\sim 10^5 \; \rm K$, derived from Eq. \ref{eq:gravpot}), and as our SNe are injected just as thermal energy, there is no characteristic temperature for this gas. Since all of the hot gas in our simulations has been produced by the action of SNe it is reasonable to suppose that the temperature of this phase may be determined by the transition from the adiabatic to the momentum driven phases, as described by \citet{Cox_72, Chevalier_1974} and \citet{Larson_1974}. 

In this explanation, the supernovae would rapidly expand in the adiabatic phase until the action of cooling relative to expansion causes the growth of the remnant to decelerate, and the edge to form a cold dense shell. This shell still expands, but at a considerably reduced rate, driven primarily by the momentum of the shell. We expect the adiabatic phase to remain approximately spherical due to the short sound crossing time within the hot volume, however when the blast enters the momentum driven phase, the cooling shell is unstable and the remnant can become quite asymmetric. If the edge of the remnant reaches other sparse material the hot interior of the remnant can leak out (i.e. a `chimney' such as those seen in \citealp{Ceverino_2009}), otherwise the hot material will gradually be consumed into the dense shell as it radiates away its pressure support.

The post shock temperature, $T_s$, of the hot remnant at which the \lq sag\rq\ occurs (when cooling dominates over adiabatic expansion) was calculated in \citet{Cox_72} as 
\begin{eqnarray}
T_s &\approx& 2.0 \times 10^6 \, \left(\frac{n}{1 \; \rm cm^{-3}} \right)^{4/11} \left(\frac{E_{\rm SN}}{10^{51}\; \rm erg } \right)^{2/11} \times \nonumber \\
&& \left(\frac{\Lambda}{10^{-22} \rm erg \, cm^3 \, s^{-1} }\right)^{6/11} \, \rm K \, \label{eq:Cox_temp} .
\end{eqnarray}
The obstacle which radiates away the energy of the SN is the warm disk gas of Fig.~\ref{fig:temp_dens_vel}. Taking a mean density of these from Fig.~\ref{fig:volfrac} 
\begin{equation}
n = 3 \; \rm cm^{-3} \, \label{eq:Cox_dens},
\end{equation}
($\rho=5\times 10^{-24}$~g~cm$^{-3}$) we expect a characteristic temperature of the remnants to be $T_{\rm hot} \approx 3 \times 10^6$~K, very close to our HIM temperature of $\sim 10^{6}\; \rm K$.

Another interesting application of Eq. (\ref{eq:Cox_temp}) is to estimate the mass heated by a single supernova before it ends the adiabatic phase. By finding the amount of mass required to absorb the thermal energy of a supernova we derive
\begin{eqnarray}
M_{\rm hot} &=& \frac{2}{3} \frac{m_{\rm p} E_{\rm SN}}{k_{\rm B} T_s} \nonumber \\
&=& 1350 \,M_\odot\,\left({T_s\over 3 \times 10^6{\rm K}}\right)^{-1}{E_{\rm SN}\over 10^{51}{\rm erg}}\,, \label{eq:mhot}
\end{eqnarray}
where we have neglected the initial thermal energy of the heated gas, the SN ejecta themselves (see also \citealp{Kahn_1975}), and assumed that none of the SN energy has yet been lost radiatively. For comparison, in the model of \cite{Efstathiou_2000}, a supernova evaporates a similar mass $M_{\rm ev}\sim 540\,M_\odot$ of cold clouds. If all this hot gas were to escape from the simulation without entraining any other material we would derive a mass loading of
\begin{eqnarray}
\beta &=& \frac{M_{\rm hot} \varepsilon_{100} }{100  \; \rm M_\odot} \nonumber\\
&\approx& 13 \varepsilon_{100} \left(\frac{n}{3 \; \rm cm^{-3}} \right)^{-4/11} \left(\frac{E_{\rm SN}}{10^{51}\; \rm erg } \right)^{9/11} \times \nonumber \\
&&\left(\frac{\Lambda}{10^{-22} \rm erg \, cm^3 \, s^{-1} }\right)^{-6/11} \label{eq:betamax}  \\
&\approx & 13 \varepsilon_{100} \left( \frac{ \Sigma_{\rm g}}{10 \, \rm M_\odot \, pc^{-2}} \right)^{-8/11} \left( \frac{ f_{\rm g}}{0.1} \right)^{4/11}  \times \nonumber \\
&& \left(\frac{E_{\rm SN}}{10^{51}\; \rm erg } \right)^{9/11} \left(\frac{\Lambda}{10^{-22} \rm erg \, cm^3 \, s^{-1} }\right)^{-6/11} \label{eq:beta_surf}\, ,
\end{eqnarray}
where in Eq.~(\ref{eq:betamax}) we have used the warm cloud density $n=3$~cm$^{-3}$ from Eq.~(\ref{eq:Cox_dens}), and in Eq.~(\ref{eq:beta_surf}) we have used the hydrostatic mid-plane density from Eq.~(\ref{eq:density}). The mass loading is higher at lower surface densities (and also volume densities), at higher gas fractions, and for gas that cools more slowly, and increases with the SN energy injected. If all the gas escapes at $T=T_s$ then this is an upper bound for the mass loss, since some energy will be converted to other forms such as radiation and turbulent motion, and for this simulation we do find the measured $\beta$ is significantly below this (see section \ref{sec:fitparams}). Notably many versions of semi-analytic models such as \galform\ assume $\beta$ close to this maximum.

In this section we have described a snapshot of a simulation of a patch of the ISM with similar parameters to that of the solar neighbourhood. We have reproduced a warm and hot phase in order-of-magnitude pressure equilibrium, with a value similar to that estimated for the local volume. We have explored the relation between the temperature of the hot phase and related this to the density of the warm phase via the energy of each SN and the cooling time of the gas. 

\subsection{Time dependence}
We now turn our attention to the time dependence within our simulation. We have seen in Fig.~\ref{fig:temp_dens_vel} that our idealised disk is disrupted by the energy injection from supernovae, and we are interested in the evolution that results from this. The injected energy can be converted into a number of forms, heating of the warm phase, the thermal energy of the hot phase, the mechanical energy of turbulence and the wind, the gravitational potential of the gas as it is lifted out of the disk, and the photons lost through radiative cooling. It is worth recalling that cooling is one of two ways in which energy can leave the simulation volume, the second being the advection of mass across the vertical boundaries of the simulation, taking with it the thermal, mechanical and gravitational potential energy of the gas.

\begin{figure}
\centering
\includegraphics[width=\columnwidth]{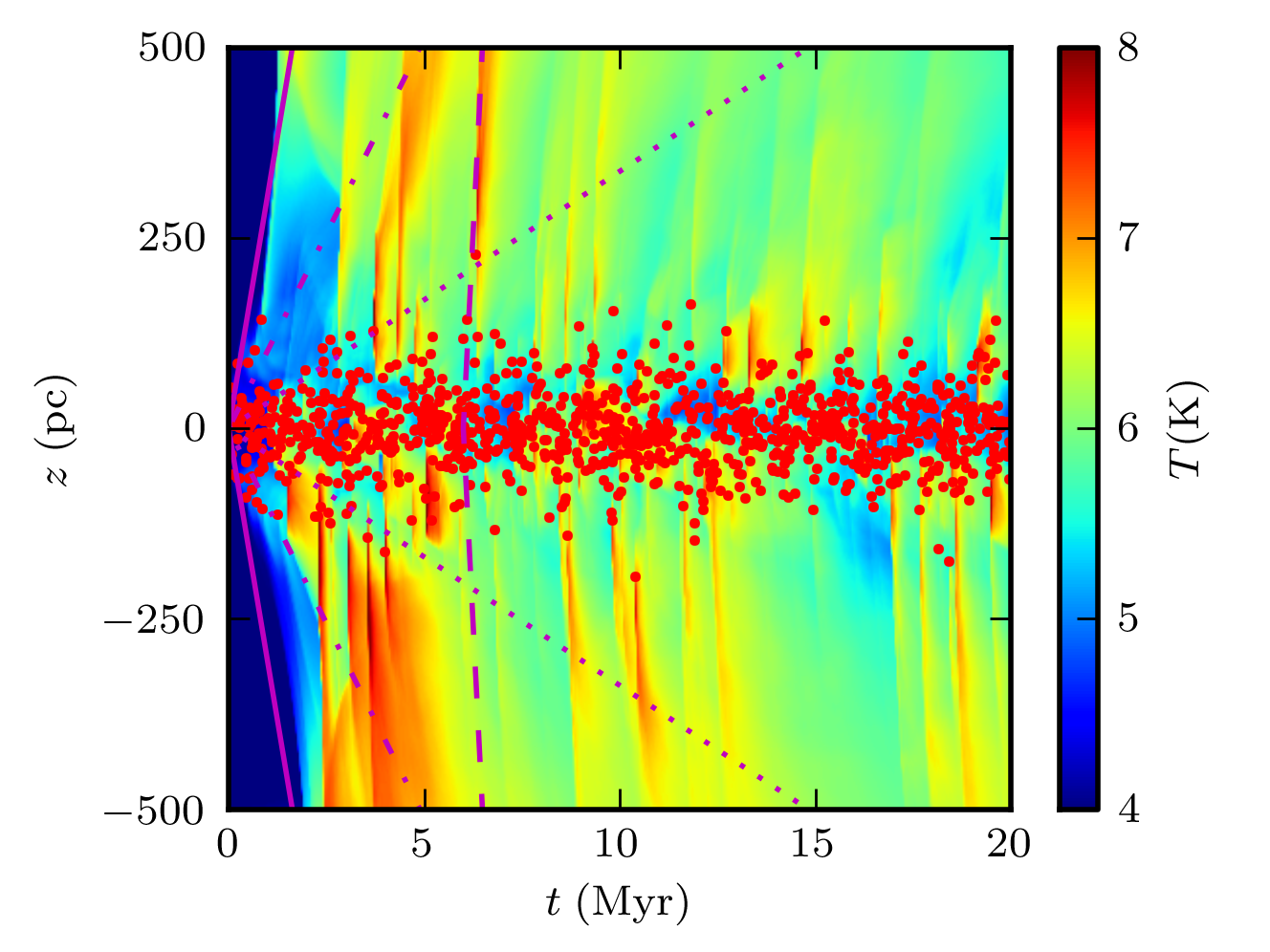}
\caption[Volume weighted mean temperature as a function of height and time, for the for the fiducial disk parameters]{Volume weighted mean temperature as a function of height and time, for the for the fiducial disk parameters yet in a wider box of $800\times 800 \times 1000$pc. At each height we have taken the average over a horizontal slice. Superimposed are \emph{red dots} indicating the locations and times of the SN events. As the simulation progresses the activity of many SNe shock heat gas and drive a vertical wind from the disk at around $300 \rm km\, s^{-1}$. \emph{Dotted, dot-dashed, solid} and \emph{dashed} magenta lines denote outflows of 33, 100, 300 and 1000 $\rm km\, s^{-1}$ respectively. Subsequent and around each supernovae can be seen a pulse (in orange) in the temperature. After a short ($t < 1$ Myr) flurry of supernova activity within the disk ($z=\pm 53 \rm pc$), the shocked regions begin to combine and rise out of the disk and the simulation volume. Occasionally, individual supernovae high above the disk (where the gas density is low) make a significant individual contribution to the wind. The 1000 $\rm km\, s^{-1}$ line has been offset to start at 6 Myr to be compared with the propagation of one of such temperature pulses.}
\label{fig:height_time}
\end{figure}

Fig.~\ref{fig:height_time} is a `space-time' plot of the onset of the outflow: time is along the horizontal axis, and the projected mean temperature, $\bar T$, as a function of height is colour coded and shown on the vertical axis, red dots correspond to the times and location of individual SN injection events. In order to reduce the effects of stochastic outflows we performed this simulation in a larger box, of width $800$ pc.
The initially hydrostatic gas at temperature $T=T_0$ seen at the far left of the figure is quickly replaced by gas at a range of temperatures.
The dark blue coloured band, corresponding to $\bar T\approx T_0$, episodically widens as a function of time, as the disk puffs up.
Gas with a mean temperature $\bar T\sim 10^6$~K is seen to stream out of the disk at a range of velocities. From Fig.~\ref{fig:volfrac} we recall that there is actually very little gas by mass at $10^6$~K, however by volume the mean temperature will be close to this.  Around each supernova a plume of hot gas can be seen (cyan against the colder dark blue gas). At late times these plumes combine and drive the galactic wind.

Comparing with the velocity lines we can see the evolution of the outflow velocity with time, with many structures with velocities in the range of 30-300$\;\rm km\, s^{-1}$. Superposed, however, are some extremely steep (w.r.t. time, i.e. high velocity) discontinuities where much of the simulation volume rapidly experiences an increase in temperature. These appear to propagate from individual SNe, and race away from the disk with velocities in excess of $1000\; \rm km\,s^{-1}$,  consistent with a sudden pressurisation of the hot phase of the ISM\footnote{For reference, the temperature that correspond to a given sound speed $c$ is $T=7.3\times 10^7\,{\rm K}\,(c/1000~{\rm km}~{\rm s}^{-1})^2$.} This increased pressure causes stripping from the warm material as shocks drive in to the warmer region of the cloudy medium, adding to the mass of the hot phase.

To analyse our simulations we reduced our data set down to the following parameters, listed below. These are chosen to give us a broad overview of the evolution of the star forming disk, rather than information on the individual cells and clouds. For these parameters there is some freedom of definition, e.g. when one attempts to measure the pressure one could take the mid-plane pressure, the pressure within the star forming scale height $b$, the mean pressure within the simulation volume, or the mean pressure within a volume adjusted by some measure of the current disk scale height. In all cases we have attempted to choose a definition which strikes the balance between reducing stochasticity (some candidate measures show considerably more noise than others) and ease of physical intuition.
\begin{enumerate}
\item{Mass ejection}, $\Sigma_{\rm ej}(t)$, is the amount of gas ejected from the disk per unit area. This is calculated from the mass advected through the boundary at $z = \pm 500 \rm pc$, divided by the surface area of the simulated column. This quantity is used in the calculation of the cosmologically important quantity $\beta = \dot{\Sigma}_{\rm ej} / \dot{\Sigma}_\star$ where we have identified the mass ejected from the idealised disk with the mass ejected from the galaxy. To achieve the nearest correspondence we try to maximise the volume we are measuring the loss from, i.e. the entire simulation volume. The corresponding normalised quantity is the fraction of gas remaining in the disk, $f_\Sigma \equiv 1 - \Sigma_{\rm ej} / \Sigma_{\rm g}$.

\item{Cold gas/Hot gas surface density} is the remaining cold/hot gas surface densities in the simulation volume, and in combination with the mass ejected, sum to the initial gas surface density $\Sigma_{\rm g}$.

\item{Cold volume fraction}, $f_{\rm cold}$, is the volume fraction of cold gas, sometimes quoted in terms of the porosity 
\begin{equation}\label{eq:porosity}
P = - \log f_{\rm cold}\,,
\end{equation}
\citep{Silk_2001}. We distinguish between cold and hot phases at a cut-off of $2T_0$ (i.e. twice the lower limit of our cooling function). Though the choice of $2T_0$ may seem arbitrary, it is apparent from Fig.~\ref{fig:volfrac} that the bi-modality of the warm and hot phases is quite strong, so the dependence of our results on the choice of temperature cut-off is rather low. Since the effectiveness of SNe in driving feedback is highly suppressed in dense (and cold) regions, the volume filling factor largely determines the probability that an individual supernova will explode in the hot phase.  The volume we study is $z \in [-250,250] \; \rm pc$, as we are not interested in the hot gas far from the plane of the disk (where SNe do not occur). 

\item{Pressure}, $p$, is the mean pressure in the entire simulation volume. Hot material from the disk is ejected by a mean pressure gradient to the edge of the simulation volume, however the stochastic nature of supernova events creates a significant variation over small time scales and large spatial scales\footnote{The pressure equilibrium predicted by \cite{Spitzer_1956} holds over smaller spatial scales where the supersonic turbulence decays over the sound crossing time.} and thus it is desirable to smooth the pressure estimate over as large a volume as possible.

\item{Half-mass height}, $\lambda_{1/2}$, is defined as the height where $z \in [-\lambda_{1/2}, \lambda_{1/2}]$  contains half the original gas mass of the disk,
\begin{equation}
\label{eq:scale_height}
\lambda_{1/2} = \min \left\{ z' : \int_{-z'}^{z'} \left< \rho \right>_z {\rm dz}  > \frac{1}{2} \Sigma_{\rm g} \right\}\,.
\end{equation}
At the start of the simulation this is related to the scale height by our choice of isothermal density profile, at $\lambda_{1/2} = \frac{1}{2} b \log 3$. Large outflows will `puff-up' the disk to greater scale heights, at late times this would become inconsistent with our star formation profile.

\item{Effective cooling rate}, $\eta_{\rm eff}$, is the total radiative cooling rate in the simulation volume divided by the mean SNe energy injection rate,
\begin{equation}
\eta_{\rm eff} = \frac{\int_V \Lambda n^2 {\rm dV}}{\int_{\rm area} E_{\rm SN} \epsilon_{100} (\dot{\Sigma}_\star / 100 {\rm M_\odot}){\rm dA}} \, .
\end{equation}
Conservation of energy implies that all of the energy not released as radiation must end up either in the wind or as gravitational potential energy. Due to the discrete nature of time sampling with snapshots (i.e. for many of the quantities such as cooling and we have instantaneous measurements of their time derivatives and not measurements of the integrated quantities themselves) there is some error on our estimate of the integrated quantities. Most susceptible is the estimate of the cooling rate: the tail of high-density gas seen in the density probability distribution function of Fig.~\ref{fig:volfrac}, cools very rapidly, and our time sampling means its contribution to cooling is under-estimated. We will inevitably miss some cooling that would have occurred outside the simulation volume (although much of this gas is tenuous and will have a long cooling time, little gas remains dense in the outflowing material). Nevertheless our high snapshot frequency run gives us energy conservation to $\sim 1 \%$ and confidence that we can accurately measure the outflowing components from the low frequency runs (energy conservation in the simulation itself is of course much better than this.)

\end{enumerate}

\begin{figure}
\centering
\includegraphics[width=\columnwidth]{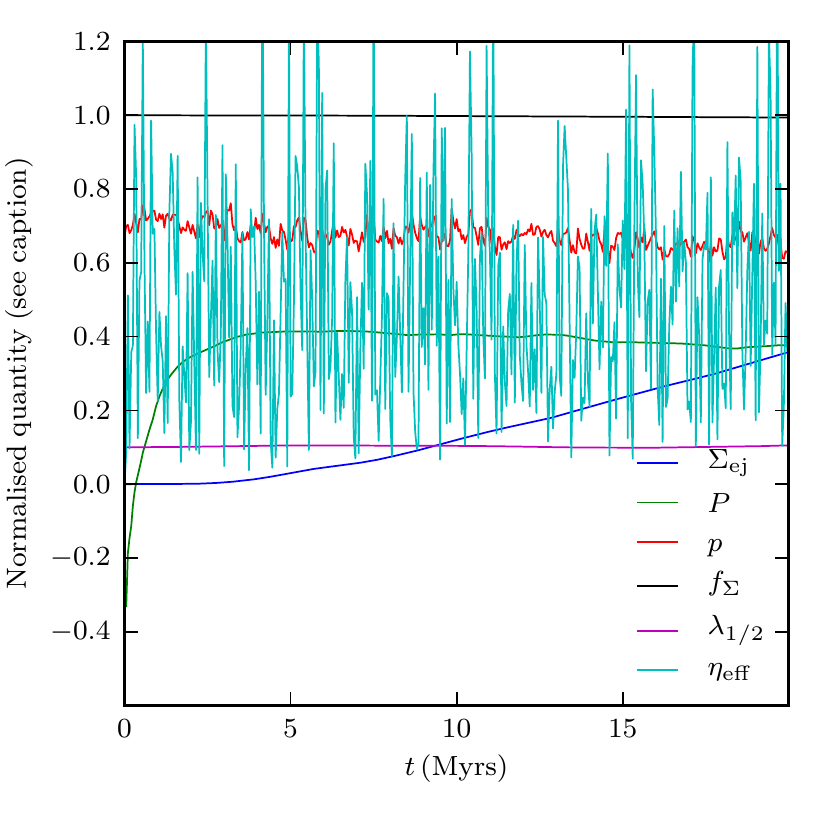}
\caption[Time evolution of statistics for the run in Fig. \ref{fig:height_time}]{Generation of an outflow in the run in Fig. \ref{fig:height_time} as characterised by the evolution of normalised quantities described in (i)-(v) in the text. After a transient initial stage of $\sim 5$~Myr, gas starts to be ejected at a nearly constant rate of $\sim 0.01\,M_\odot\,{\rm Myr}^{-1}\,{\rm pc}^{-2}$.
The \emph{dark blue line} is the cumulative mass ejected per unit area, in units of $0.2 \rm M_\odot \, \rm pc^{-2}$. The porosity $P=-\log(f_{\rm cold})$ of hot gas builds very quickly, \emph{green line} is $0.5+0.2P$, implying a filling factor of the HIM of approximatelty 50\%. The \emph{red line} is the mean pressure, $\log_{10} \left( p/ 10^{3} {\rm K \, cm^{-3}} \right)$, disturbed from its initial value of $0.7\times 10^3$~K~cm$^{-3}$ in the base of the disk by action of the SNe. \emph{Black line} is the fraction of gas remaining in the simulation. The \emph{magenta line} is  the evolution of the scale height, Eq.~(\ref{eq:scale_height}),  in terms of $0.1 \lambda_{1/2}(t) / \lambda_{1/2}(0)$. The highly stochastic \emph{cyan line} is $\eta_{\rm eff}$, the instantaneous cooling rate as a fraction of the mean SNe energy injection rate.   During the first $\sim 2$ Myr the porosity in the simulation rapidly increases, after which the material begins to be ejected from the simulation in a relatively linear fashion.  There are periods where the cooling rate increases dramatically by a factor $\sim 10$,  which are closely related to SN energy injection events.  Energy injection has not significantly puffed up the disk.}
\label{fig:single_evolution}
\end{figure}

In Figure \ref{fig:single_evolution} we inspect these parameters for the simulation in Fig. \ref{fig:height_time}. For the first $\sim 2$ Myr, the most notable feature is the rapid increase of porosity as the supernova blasts evacuate bubbles in the disk. The height of the disk remains approximately constant. As the simulation evolves, the remaining gas fraction declines (black curve) as gas leaves the simulation volume (blue curve). The mass lost from the simulation appears to be a nearly linear function of time at this stage, suggesting a constant outflow rate, which we investigate further in section \ref{sec:fitparams}.

\subsection{Comparison to a rarefaction zone}\label{sec:rarefaction}
A characteristic feature of both simulated and observed outflows \citep{Steidel_2010} is that the wind speed {\em increases} with height $z$ above the disk, and it has been suggested that radiation driving is the cause of this \citep{Murray_2005}. Since radiation driving is not included in our modelling yet the outflow does accelerate, we suggest the following physical model. The combined effects of several supernova explosions cause the ISM pressure to increase substantially above the hydrostatic equilibrium value. If gravity is not dominant, this will lead to the higher pressure ISM expanding into the lower pressure regions above the disk. In the launch region of such an outflow, 1D (plane-parallel) symmetry is a reasonable description of the geometry. A useful comparison is the behaviour of a rarefaction wave, where a homogeneous static gas is released into a sparse, pressure free zone, and for which the similarity solution is 
\begin{eqnarray}
v(\eta) &=& \frac{2}{\gamma + 1} c_0\,\left( 1 + \eta \right) \nonumber\\
\rho(\eta)  &=& \rho_0 \left( {2\over \gamma+1}- {\gamma-1\over\gamma+1}\,\eta \right)^{2/(\gamma-1)} \nonumber\\
\eta &\equiv& {z\over c_0 t}\,,
\end{eqnarray}
valid for
\begin{equation}
\eta\in \left[ -1, \frac{2}{\gamma - 1}\right]\,.
\end{equation}
In such a flow, speed increases with height $z$ and density decreases. This is distinct from the flow due to a single blast wave, since in the Sedov-Taylor phase density {\em increases} with distance from the blast, which is not the case for the disk outflow (Fig.~1). Notably this does not describe a \emph{steady} wind, which would be the result of continuous energy injection.

\begin{figure}
\centering
\includegraphics[width=\columnwidth]{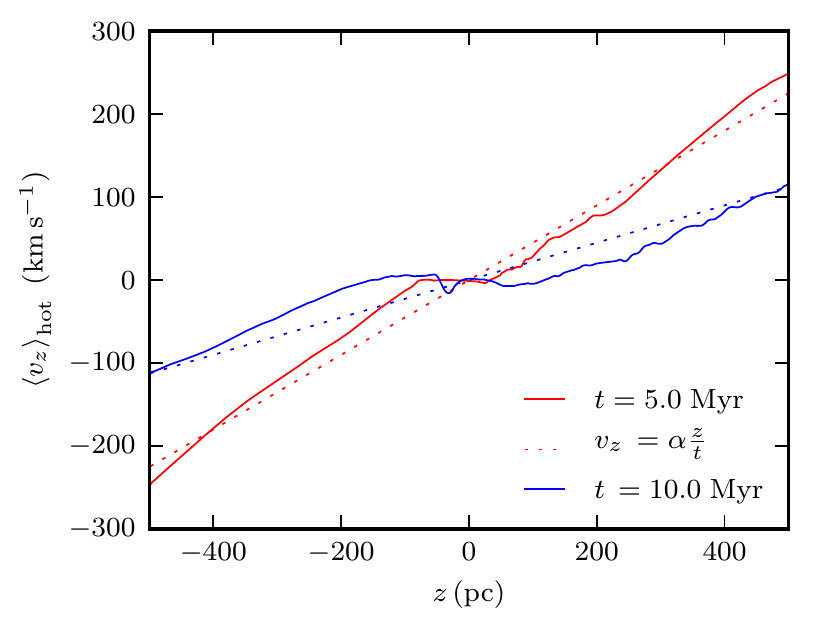}
\caption{\emph{Solid line} shows the mean vertical velocity as a function of height for two times in the $\Sigma_{\rm g} = 2.5\; \rm M_\odot\, pc^{-2}$, $f_{\rm g}=0.01$ simulation showing only the hot gas (where we have defined hot gas to be that above $2\times 10^4 \, \rm K$).  \emph{Red dotted line} is a linear fit ($\alpha=2.6$) to the earlier snapshot ($t=2.5 \, \rm Myr$) which is then extrapolated to the later snapshot (\emph{blue dotted line}). This shows the profile is evolving in an approximately self-similar fashion with the hot material accelerating away from the disk primarily due to its thermal energy being converted to kinetic energy.}
\label{fig:rarefaction}
\end{figure}

In a rarefaction wave, the acceleration is due to the pressure gradient in the outflow, and results in thermal energy being converted to kinetic energy, and the asymptotic flow speed is $v_{\rm max}=3c_0$ for $\gamma=5/3$.  The outflowing gas above the disk is mainly warm ISM gas that is entrained by the hot SN bubbles that power the rarefaction wave. Figure~\ref{fig:rarefaction} shows the behaviour of the simulation to be consistent with this model: velocity increases with height $z$, but decreases with time at a given height in way predicted by the similarity solution.  

Notably the rarefaction is not a steady-state solution, and thus is not a good description of the time-averaged behaviour of the gas. Such behaviour should mimic the result of continuous energy injection, where multiple overlapping SNe in the form of rarefactions or Sedov-Taylor blast waves (see e.g.  \citealp{Castor_1975, Weaver_1977, McCray_1987}) drive a large-scale wind. Our simulations are sufficiently stochastic however that we shall leave this for future work. There will also be departures from a steady state solution as the disk consumes its gas, or in a real galaxy, has some gas inflow.

\begin{figure*}
\centering
\includegraphics[width=2\columnwidth]{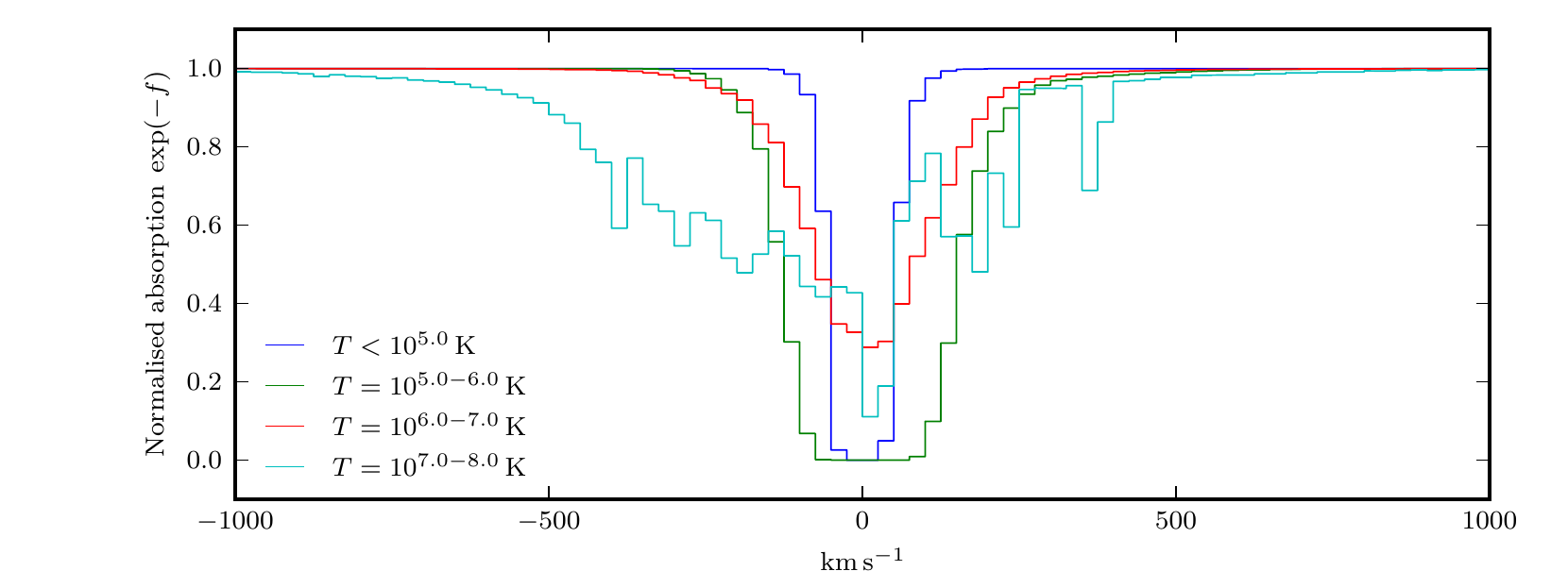}
\caption{Normalised column density as a function of velocity, for gas with different different temperature (\emph{coloured lines}). For low temperature absorbers ($\lesssim 10^6 \; \rm K$) we to see a single peaked profile centred around the rest frame velocity of the disk. For higher temperatures absorbers, we see absorption at higher velocities relative to the disk, with velocity increasing with temperature.  Only the $\gtrsim 10^{7} \; \rm K$ distribution appears to show any significant asymmetry.}
\label{fig:outflow}
\end{figure*}

\subsection{Absorption features of galactic winds}\label{sec:MockAbsorb}
\citet{Steidel_2010} proposes that the C{\sc II} absorption line data is also well fit with velocities increasing with distance from the disk (in particular the lower panel of Fig. 24 of \citealp{Steidel_2010}). The explanation above provides a physical mechanism for those measured features. This is without the radiation and dust driven mechanisms invoked by \citet{Murray_2005, Martin_2005, Sharma_2011}.

We pointed-out in Fig.~\ref{fig:temp_dens_vel} the multi-phase nature of the outflow, as well as the fact that outflow speed depends on temperature.
This is made more vivid in Fig.~\ref{fig:outflow} in which we show mock \lq absorption lines\rq\ of gas selected in narrow temperature bins. These mock line profiles are simply the fraction of gas in a given temperature range, that is moving with a given velocity, as a function of velocity, $v$. For the temperatures $T< 10^7$~K, the lines have their highest optical depths at $v\sim 0$~km~s$^{-1}$, and shapes which with vary little with temperature, $T$, and are almost symmetric in velocity. The line shapes broaden as the temperature increases, and for the hottest gas at $T>10^7$~K the line becomes asymmetric and the absorption centre is now $\sim -100$~km~s$^{-1}$. It is tempting to compare these to absorption line studies in outflows such as \cite{Martin_2005} in NaI and \cite{Weiner_2009} in MgII, however more work would be required to calculate corrections for the geometry and ionisation.

Fig.~\ref{fig:temp_dens_vel} also shows colder clouds entrained inside the much hotter flow, with cometary-like tails where the cloudy medium is being ablated by the hot gas rushing past. Absorption lines might arise from mass loading this hot flow either through conductive evaporation (see for example \citealp{Boehringer_1987,Gnat_2010}) and/or through ablation (e.g. \citealp{Hartquist_1986}). \cite{Fujita_2009} investigated the warm clouds in axisymmetric 2-dimensional simulations, where the clouds appear as Rayleigh-Taylor unstable cool shells and fragments that can explain the high velocity Na I absorption lines. We note that the metallicity of the gas phases is likely to be quite distinct, as the supernovae are both the origin of the heating and of the metals, and we intend to explore this in a subsequent paper.

\section{The dependence of outflows on disk properties}\label{sec:fitparams}
\label{sect:statistics}
In the previous section we have discussed in detail the features of a simulation of a supernova-driven wind using a set of fiducial parameters for the disk and supernova rate, the processes which drive it and the statistics that can be used to examine it. In this section we explore how the outflow properties 
vary and scale with the parameters. We will use such scalings in the next section to integrate over a full galactic disk.

\begin{figure}
\centering
\includegraphics[width=\columnwidth]{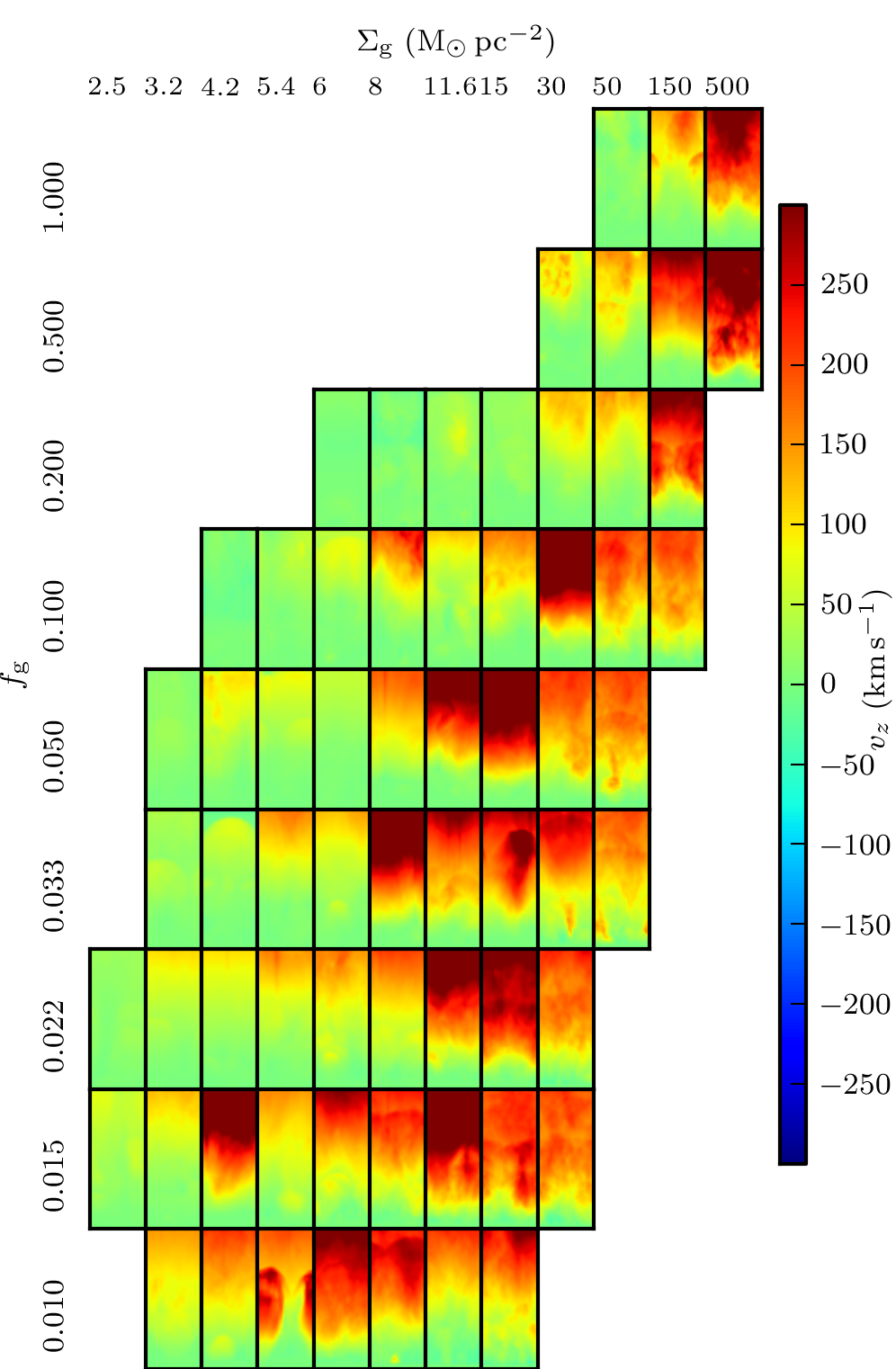}
\caption[Matrix of upper half plane velocities for different simulations]{Matrix view of simulations varying gas surface density ($\Sigma_{\rm g}$) and gas fraction ($f_{\rm g}$), each panel showing a time averaged vertical velocity for the upper half plane of each simulation (i.e. the disk is at the base of each panel. Gas surface density increases from left to right, gas fraction increases from bottom to top. There appears to be a strong trend in wind velocity towards the lower right panels, i.e. a disk with low gas fraction but high gas surface density tends to generate a faster wind.}
\label{fig:matrix_pics}
\end{figure}

In Fig.~\ref{fig:matrix_pics} we plot average velocities above the simulation disk, for the simulations varying $\Sigma_{\rm g}$ and $f_{\rm g}$. There appears to be a strong trend in wind velocity, with wind speed increasing with increasing gas surface density, but decreasing gas fraction. There are no simulations in the upper left as these would have a scale height larger than half the box size, or in the lower right as these would have a scale height less than $3$ pc.

\subsection{Mass outflow}
Inspecting the ratio of mass outflow rate to star formation rate gives us an analogous property to that of Eq. (\ref{eq:beta}), i.e. for a specific area on the disk
\begin{equation}
\beta = \frac{\dot{\Sigma}_{\rm ej}}{\dot{\Sigma}_\star} \, ,
\end{equation}
which we use in our subsequent analysis. In theory every snapshot from our simulations contains an estimate of this $\beta$, as the mass outflow rate at a specific height, however this is rather stochastic, and as an alternative we calculate $\beta$ as a fit to several measurements of the integrated outflow
\begin{equation}
y_i = \frac{\int_0^{t_i} \dot{\Sigma}_{\rm ej} {\rm d}t }{\int_0^{t_i} \dot{\Sigma}_\star {\rm d}t} \, ,
\end{equation}
which are easily obtained from each simulation snapshot.  We fit the data samples $\left\{(t_i, y_i)\right\}_{i=1}^n$ with the ramp function, 
\begin{equation} \label{eq:betafit}
f(t) = \left\{ \begin{array}{cc} 
 0 , & t< t_0 \\
\beta t ,&  t \geq t_0\,,
 \end{array} \right.
\end{equation}
where the parameters $t_0$ and $\beta$ are free variables. The motivation for choosing such a fit is that, whilst the ejection rate is nearly linear in most cases, there is a time ($t_0$) required for the system to reach a quasi steady state. This will not be a true steady state, in that the wind will eventually exhaust the supply of cold gas, however this occurs over a sufficiently long time-scale that the fit is a reasonable description for our simulations.

The square error of this function can be analytically solved by finding linear regressions for the subsets $s_k$ of $\left\{ (t_i,y_i) \right\}_{i=1}^n$ defined by $\left\{ (t_i,y_i) \right\}_{i=k}^n$ and choosing the minimum $k$ such that the linear regression $t$-intercept $<t_k$. If we define $g(s_k)$ as the $t$-intercept of the linear regression for $s_k$, then
\begin{equation}\label{eq:t0}
t_0 = \min \left\{ t_k : g\left(s_k\right)<t_k, s_k \equiv \left\{ (t_i,y_i) \right\}_{i=k}^n \right\} \,,
\end{equation}
and $\beta$ is the slope of this linear regression.

Plots of the gas fraction remaining in the simulation volumes can be seen in Fig. \ref{fig:single_evolution} for the fiducial model, and for the set of simulations of varying $\Sigma_{\rm g}$ and $f_{\rm g}$ in Fig. \ref{fig:sliceA} in the Appendix where we also show the fits given by Eq. (\ref{eq:betafit}).

\begin{figure}
\centering
\includegraphics[width=\columnwidth]{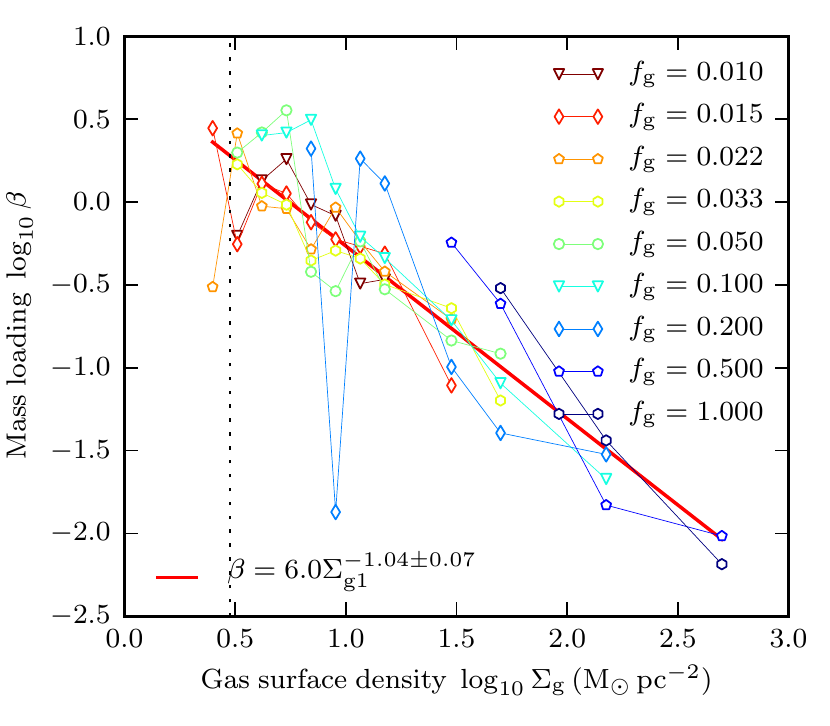}
\caption{The mass loading $\beta$ (mass ejection rate vs. rate of star formation) as a function of gas surface density $\Sigma_{\rm g}$. Each point represents a fit of $\beta$ (section \ref{sec:fitparams}) to a star formation simulations of varying $\Sigma_{\rm g}$ and $f_{\rm g}$. \emph{Red line} denotes a power law fit with jack-knife errors, \emph{coloured symbols} (red-blue) correspond to the simulations with gas fraction $f_{\rm g} = $ 0.01 - 1.0 respectively. \emph{Vertical grey dashed line} indicates the $3\; \rm M_\odot \, pc^{-2}$ threshold for star formation from \citet{Schaye_2004}. We see a significant negative dependency of $\beta \sim \left( \Sigma_{\rm g}/ 1\; {\rm M_\odot \, pc^{-2}} \right)^{-1.04\pm 0.07}$ on the gas surface density, which may be due to the larger gravitational potential or the higher rate of cooling (incurred by higher gas densities) or some combination of both. We also note that the scatter seems partially a function of $f_{\rm g}$, with higher gas fractions showing larger $\beta$'s than the lower (e.g. \emph{blue} vs. \emph{green}).}
\label{fig:betafits}
\end{figure}

In Fig. \ref{fig:betafits} we plot the mass loading $\beta$ as a function of gas surface density $\Sigma_{\rm g}$. Each point represents a fit of $\beta(\Sigma_{\rm g})$ for the simulations varying $\Sigma_{\rm g}$ and $f_{\rm g}$. The first point to note is that our $\beta$ values all lie below $4$, and for a large range of our parameters $\beta \ll 1$, i.e. our domain of parameter space switches from effective feedback (more gas ejected than stars formed) to ineffective, where the amount of gas released is much smaller than that converted into stars.

Based on jack-knife errors, our power law fit shows a significant negative dependency, $\beta\approx 6 \left( \Sigma_{\rm g}/ 1\; {\rm M_\odot \, pc^{-2}} \right)^{-1.04\pm 0.07}$, implying that at high gas surface densities the feedback is less efficient. This could be due to a number of effects. Since a higher gas surface density will correspond to a deeper potential well, the escape velocity of the gas is higher. Secondly, the higher gaseous surface densities correspond to higher gas volume densities (Eq.~(\ref{eq:density})), resulting in shorter cooling times. 

\begin{figure}
\centering
\includegraphics[width=\columnwidth]{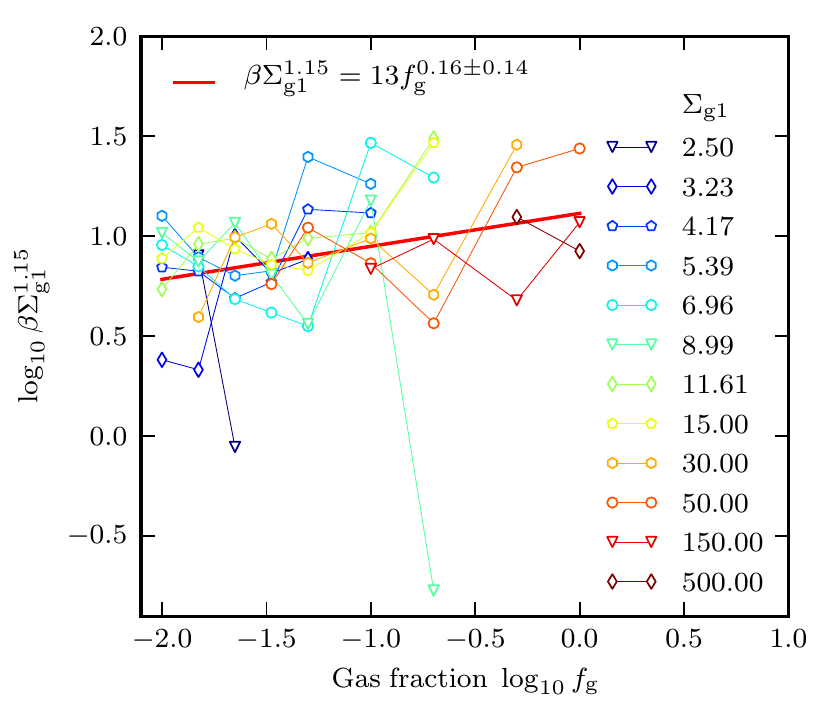}
\caption{Joint dependence of the mass loading $\beta$ on gas surface density, $\Sigma_{\rm g}$, and gas fraction $f_{\rm g}$. Differently coloured curves correspond to simulations with different values of $\Sigma_{\rm g1}\equiv \Sigma_{\rm g} /{\rm M}_\odot\,{\rm pc}^{-2}$, the \emph{thick red line} is our best fit of the
simulation points.  We see a dependence of $\beta \Sigma_{\rm g}^{\muEst}$ on gas fraction, with a power law dependency of $\nuEst \pm 0.15$. Higher gas fractions for a given gas surface density imply a shallower potential well, explaining why the outflow efficiency increases with $f_{\rm g}$.}
\label{fig:betafgasfits}
\end{figure}

Another notable dependency is that on the gas fraction. Some of the scatter seen in Fig. \ref{fig:betafits} actually depends systematically on the gas fraction, $f_{\rm g}$, with higher gas fractions showing consistently larger $\beta$'s than the lower values. We explore this in Figure \ref{fig:betafgasfits}, where we have performed a simultaneous fit of $\beta$ to both the gas surface density and the gas fraction,

\begin{equation}\label{eq:log_linear_fit}
\beta = \beta_0 \Sigma_{\rm g1}^{-\mu} f_{\rm g}^\nu \, ,
\end{equation}
where we find the values
\begin{eqnarray}
\beta_0 &=& \betaEst \pm 10 \label{eq:beta_fit}\\
\mu &=& \muEst \pm 0.12 \label{eq:mu_fit} \\
\nu &=& \nuEst \pm 0.14\,, \label{eq:nu_fit}
\end{eqnarray}
By construction the joint fit now no longer shows a systematic dependence on either $\Sigma_{\rm g}$ or $f_{\rm g}$.

Accounting for this shows a positive dependency of $f_{\rm g}^{\nuEst \pm 0.14}$, i.e. by holding the gas surface density constant but increasing the gas fraction (which reduces the gravitational strength, thus increasing the dynamical time and reducing the star formation rate) increases the mass loading. As with the dependence on gas surface density, we are effectively seeing a sub-linear dependence on star formation rate, as we decrease the star formation (increase the gas fraction), we see a less than proportionate drop in the outflow rate. Again, the mechanism causing this should be a combination of the processes for the $\Sigma_{\rm g}$ dependence, derived above.

In Fig.~\ref{fig:betafgasfits} there is considerable scatter, especially at high gas fraction where a number of simulations have mass ejection rates considerably above the trend. This is most likely due to heavy disruption of the disk out of the plane where the wind from subsequent supernovae can eject it from the simulation volume. With such stochasticity the description of all the simulations with a simple power law becomes inadequate.

Our measured value for the exponents $\mu=\muEst$ and $\nu=\nuEst$ that relate mass loading to gas surface density and gas fraction, $\beta\propto \Sigma_{\rm g}^{-\mu}\,f_{\rm g}^\nu$, can be compared with the values from the model described in Section \ref{sec:CharacTemp}, which predicts scalings of $\mu=8/11=0.72$ and $\nu=4/11=0.37$. That model does not include gravity, and we suggest this is why the measured and predicted values differ. To verify this we have performed a series of simulations with significantly higher star formation rate, described in the Appendix. This uses a slightly different parameterisation that is more commonly used in cosmological simulations which introduces an extra dependence on the gas fraction, but the primary effect is an increase in star formation for the parameter range we study. In these runs, the energy injection rate is much higher, the volume filling factor of the hot phase much larger, and the outflow rates are correspondingly larger as well. Consequently the effect of gravity of the disk is much reduced.  Fitting $\beta\propto \Sigma_{\rm g}^{-\mu}\,f_{\rm g}^\nu$ to these runs yields $\mu=0.82$ and $\nu=0.48$,  in much better agreement with the predictions of the simple model.

It would be interesting to extend the model to account for the gravity of the disk, along the lines
followed by \cite{Stringer_2011}. Assume that the $\beta$ of the hot gas in Eq.(\ref{eq:beta_surf}) is modified by an escape fraction $f_{\rm esc}$, which is equal  to the fraction of material that has a temperature above the escape temperature of the simulation volume. Assuming the outflow has a range of temperatures, characterised by a Maxwell-Boltzmann distribution, and that only gas with $T>T_{\rm esc}$ escapes, the fraction is
\begin{eqnarray}
f_{\rm esc} &=& \int_{T_{\rm esc}} f(T) {\rm d}T \\
& \approx & 1 - \frac{4}{3 \sqrt{\pi}} \left( \frac{T_{\rm esc}}{T_{\rm s}} \right)^{3/2} \, .
\end{eqnarray}
We have assumed that $T_{\rm esc} \ll T_{\rm s}$, i.e. the low energy tail of the distribution fails to escape.  The net outflow will thus drop faster at high $\Sigma_{\rm g}\propto T_{\rm esc}$, making the dependence of the mass-loading on $\Sigma_{\rm g}$ stronger, which is consistent with the higher $\mu \approx \muEst$ we see in the lower SFR simulations.

\subsection{Radiative efficiency and energy partition in the ISM}

Whilst the mass loading of the galactic wind is one of the most cosmologically significant parameters to study, we would also like to evaluate the energy budgets and structure of the winds in our simulations. The energy injected by the SNe is absorbed into the gravitational binding energy, distributed into thermal and mechanical energy (both in the bulk motion of the wind and in turbulence throughout the simulation volume) and released as radiation (via cooling).

The energy partition also enables us to evaluate a wind velocity for the galaxy, which is commonly used to characterise feedback models for galaxy formation (e.g. \citealp{Bower_2011}). The fraction of the energy that is incorporated into the wind, in combination with the mass loading, determines the overall wind speed for a galaxy. This is an important parameter in determining whether the wind can leave the galaxy and hence provide efficient quenching of star formation.

By examining our simulations we can determine the fractions of energy that has been converted in to the different modes. In our fiducial simulation, we discover that a fraction of 87\% was radiated, 4.5\% was advected out of the computational volume as thermal energy, 5\% as mechanical energy (with over half of this in the form of turbulent energy), 1\% went into heating the simulation volume\footnote{Note that in a true steady state this fraction should be compensated by cooling.}, 1\% went into turbulence in the simulation volume and a rather low $0.5\%$ went into puffing-up the disk. The parameters here are averaged in a similar manner to the mass ejection rate, by taking the mean over snapshots after $t_0$ (Eq. \ref{eq:t0}), i.e. in the quasi-steady regime. 

Summation of these quantities allows us to estimate $\eta_T$ (Eq. \ref{eq:eta_T}), the fraction of power that is thermalised in to the outflow
\begin{equation}
\eta_T = \eta_{\rm therm} + \eta_{\rm mech} \, ,
\end{equation}
i.e. the sum of the thermal and mechanical (bulk and turbulent) contributions, (the remainder going almost entirely in to cooling). This allows us to calculate an effective velocity $v_{\rm eff}$ for the wind, 
\begin{equation}\label{eq:vwind}
v_{\rm eff} = \sqrt{\frac{2\eta_T}{\beta} \left(\frac{E_{\rm SN} \varepsilon_{100} }{100 \rm M_\odot}\right)} \, ,
\end{equation}
where we have combined the equation for mass loading, $\beta\equiv \dot M_{\rm wind}/\dot M_\star$, and the thermalisation of supernova energy into the kinetic energy of the wind ($\eta_T$), to find the specific energy in the wind (i.e an inversion of Eq.~(\ref{eq:eta_T})). Notably this will be significantly higher than the wind velocities we see at the edge of our simulation volume because it includes the energy of the thermal and turbulent components. At larger distances from the galaxy, however, we expect this to be a more realistic estimate, as the thermal energy accelerates the wind and is converted in to the mechanical energy of the bulk flow. This is a consequence of our simulations focusing on the launch region of the galactic wind, and hence the wind has not yet reached its terminal velocity.  Note that ram pressure from infalling gas may be an important obstacle in slowing down, or even preventing the outflowing gas from escaping (e.g. \citealp{Theuns_2002}).

\begin{figure}
\centering
\includegraphics[width=\columnwidth]{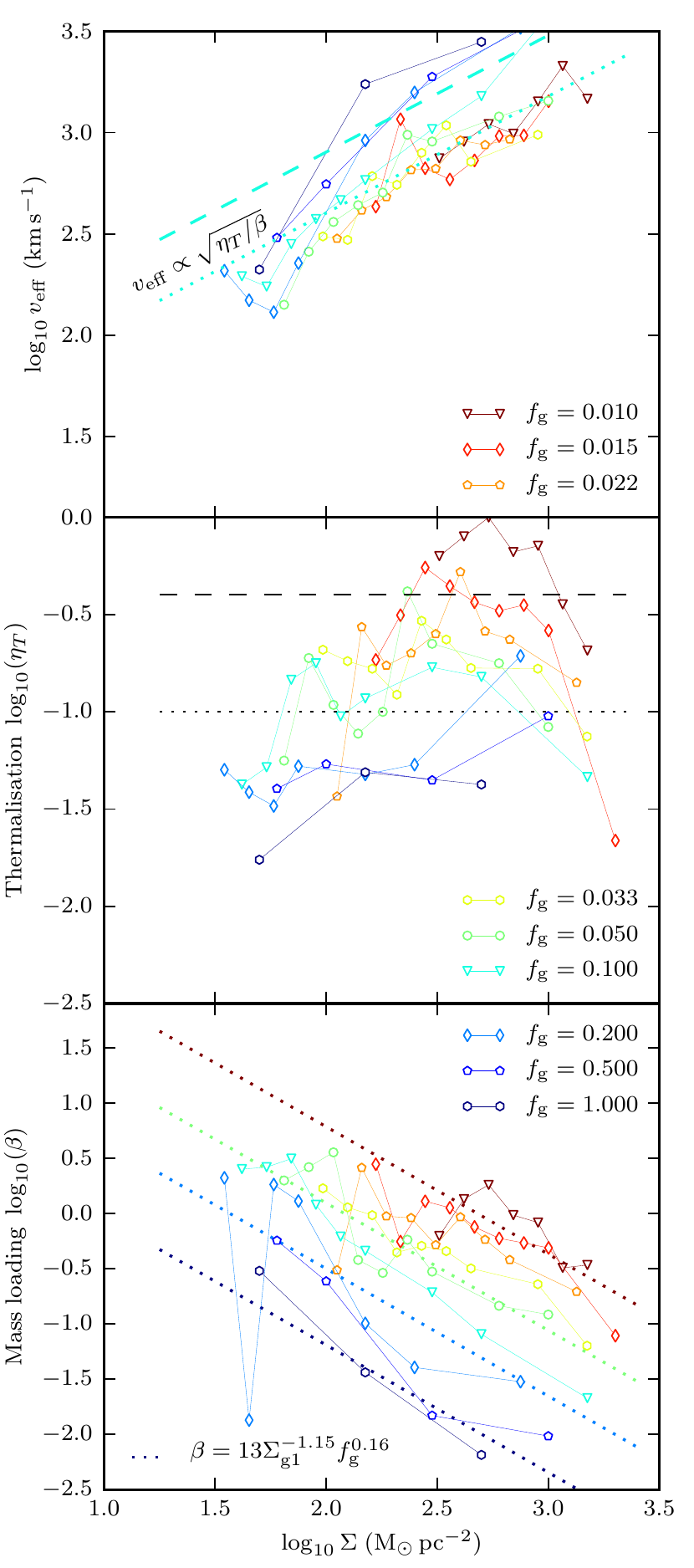}
\caption{Effective wind speed (\emph{upper panel}), outflow efficiency (\emph{middle panel}) and mass loading (\emph{lower panel}) as a function of total surface density $\Sigma=\Sigma_{\rm g}/f_{\rm g}$. Coloured lines with symbols are the simulations from figures (\ref{fig:betafits}-\ref{fig:betafgasfits}), with values of the gas fraction $f_{\rm g}$ as indicated. \emph{Dotted lines} in the lower panel are the scalings from equations (\ref{eq:beta_fit}-\ref{eq:nu_fit}), plotted for $f_{\rm g} = 0.01,0.015,0.2,1.0$ in the corresponding colours. Lines of constant efficiency, $\eta_T=0.1$ and 0.4 are shown in the middle panel (\emph{black dotted} and \emph{dashed}, respectively). Curves for the corresponding scaling of the effective wind speed for $f_{\rm g}=0.1$ are shown in the upper panel. The outflow efficiency increases with surface density, as does the effective wind speed. }
\label{fig:vwindfits}
\end{figure}

In Figure \ref{fig:vwindfits} we explore the dependence of the mass loading $\beta$, the fraction of power in the outflow, $\eta_T$, and the effective wind velocity, $v_{\rm eff},$ as a function of the total surface density of the disk, $\Sigma=\Sigma_{\rm g}/f_{\rm g}$. In terms of the mass loading we see a negative dependence on surface density, for comparison we have also included the power law fit from Eqs.~(\ref{eq:beta_fit}-\ref{eq:nu_fit}).

The fraction of power released in to the wind, $\eta_T$, appears to be correlated almost entirely with gas fraction $f_{\rm g}$, at high gas fractions much of the energy of star formation is simply radiated away, which is intuitive since the higher gas fractions will have shorter cooling times. For comparison we also show values of $\eta_T = 0.1$ and $0.4$, the former being the equivalent to the widely quoted $10\%$ efficiency in \citealp{Larson_1974}): we find star formation in disks to lie close to this value, except at very low gas fractions.

The fall in outflow power in Fig. \ref{fig:vwindfits} at low surface densities can also be seen as a fall in the effective wind velocity. Here we have converted our sample values of $\eta_T = 0.1, \; 0.4$ into effective wind velocities using the power law fit for $\beta$ in Eqs.~(\ref{eq:beta_fit}-\ref{eq:nu_fit}). Each gas fraction appears to follow a line of approximately constant $\eta_T$, although there is some suggestion of a change in slope below $\Sigma = 10^2 \; \rm M_\odot \, pc^{-2}$.

\section{Impact of outflows on galaxy evolution}
\label{sect:evolution}
In this section we apply our results from the previous section to the mass outflow from disk galaxies of different masses. We will assume a surface density profile for a galaxy and the use our fits for outflow efficiency as function of surface density, to deduce an overall feedback efficiency.

\subsection{Dependence on circular velocity from theoretical arguments}\label{sec:beta_MMW}
In this section we take our measured dependencies of the mass loading parameter (which are derived for a patch of the ISM) and apply them to an entire disk galaxy by integrating over the surface of the disk. This will allow us to compare with feedback schemes considered in \cite{Cole_2000, Bower_2006} etc., which introduce a relation between circular velocity, mass loading, and effective wind speed.

Our first step is to assume a model for a disk galaxy inside dark a matter halo where we follow \citet{Mo_Mao_and_White_98}. The circular velocity of a spherical isothermal dark halo of mass $M_{200}$ is given by
\begin{equation}
V_{200}^3 = 10 G M_{200} H(z) \label{eq:iso_halo}\,,
\end{equation}
\citep{Mo_Mao_and_White_98} where $H(z)$ is the Hubble parameter as a function of redshift, $z$. Since the baryonic component can release energy via cooling, it can collapse further to become a rotationally supported disk. Observed bulge-less disks have a near exponential profile in luminous mass of the form
\begin{equation}
\Sigma(r) = \Sigma_0 \exp \left( -r/R_{\rm d} \right)\, ,
\end{equation}
with normalisation $\Sigma_0$ and scale length $R_{\rm d}$. The mass of the disk is thus given by
\begin{equation}
M_{\rm d} = \int_0^\infty 2 \pi \Sigma(r) r {\rm d} r = 2 \pi \Sigma_0 R_{\rm d}^2 \, .
\label{eq:md}
\end{equation}

The scale length $R_{\rm d}$ is controlled by the specific angular momentum of the material forming the disk (e.g. \citealp{Fall_Efstathiou_1980}). An exponential disk with constant rotation velocity $V_{\rm d}$ has angular momentum
\begin{equation}
J_{\rm d} =  4 \pi \Sigma_0 V_{\rm d} R_{\rm d}^3 \, ,
\end{equation}
and if we parameterise in terms of the disk mass as a fraction of the halo mass, $m_{\rm d} \equiv M_{\rm d} / M_{200}$, the circular velocity of the disk as a fraction of the halo's, $v_{\rm d} = V_{\rm d} / V_{200}$ and the specific angular momentum fraction of the disk $j_{\rm d} / m_{\rm d}$, we can infer the surface density normalisation to be
\begin{eqnarray}
\Sigma_0 &=& \frac{2}{\pi} \frac{M_{\rm d}^3 V_{\rm d}^2}{J_{\rm d}^2} \nonumber \\
&=& \frac{10 H(z)}{\pi G} \lambda^{-2} \left( \frac{j_{\rm d}}{m_{\rm d}} \right)^{-2} m_{\rm d} v_{\rm d}^2 V_{200} \label{eq:MMW_sigma} \, ,
\end{eqnarray}
where $\lambda$ is the spin parameter of the isothermal halo in Eq.~(\ref{eq:iso_halo}). Notably if we set $v_{\rm d}=1$ we recover the \cite{Mo_Mao_and_White_98} surface density equation, yet for real disks $v_{\rm d} > 1$ as the contribution of baryons to the rotation velocities is not insignificant.

We can now compute a mean mass loading $\hat{\beta}$ for such a galaxy, by evaluating
\begin{equation}
\hat{\beta} \equiv  \frac{\dot{M}_{\rm wind}}{\dot{M}_\star} = \frac{\int 2 \pi \beta \dot{\Sigma}_\star r {\rm d}r}{\int 2 \pi \dot{\Sigma}_\star r {\rm d}r} \, , \label{eq:betahat_integral}
\end{equation}
where we will assume the surface density in star formation, $\dot\Sigma_\star$, follows the Kennicutt-Schmidt relation,
\begin{equation}
\dot{\Sigma}_\star = A \Sigma_{\rm g}^n \, .
\end{equation}

\begin{figure}
\centering
\includegraphics[width=\columnwidth]{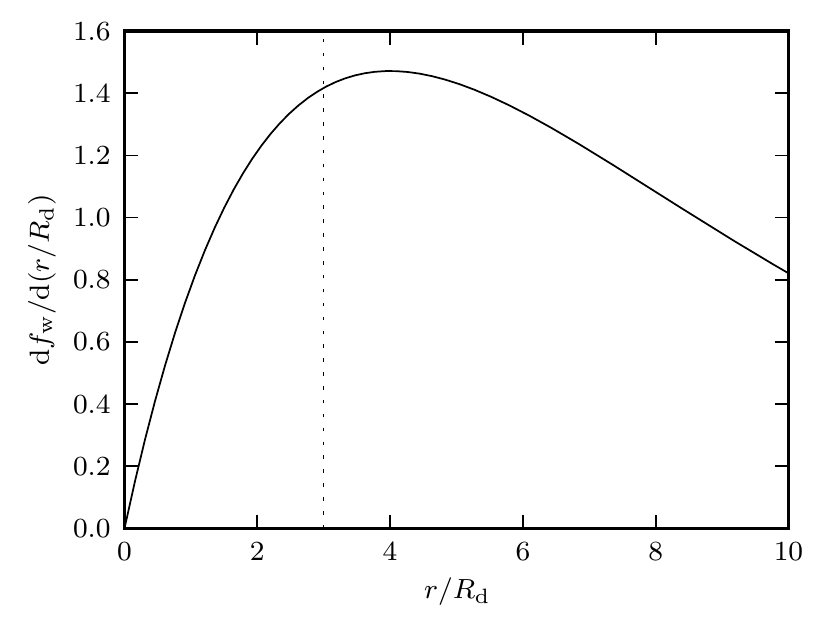}
\caption{Fraction of the wind launched at each radii in the disk (Eq.~\ref{eq:frac_wind}), for a Kennicutt-Schmidt relation $\dot\Sigma_\star \propto \Sigma_{\rm g}^n $, with  $n=1.4$, and assuming mass loading scales with gas surface density as $\beta\propto \Sigma_{\rm g}^{-\mu}$, with $\mu=\muEst$ (Eq.~\ref{eq:mu_fit}). \emph{Dotted line} indicates the characteristic wind radius $R_{\rm w}/R_{\rm d}$ for the galaxy, where the the local mass loading equals the net mass loading for the galaxy as a whole, $\hat\beta=\dot M_{\rm wind}/\dot M_\star$.}
\label{fig:disc_wind}
\end{figure}

Taking the dependence of mass loading on surface density found from our fits to the simulations, Eq.~(\ref{eq:log_linear_fit}), then Eq.~(\ref{eq:betahat_integral}) can be integrated analytically.  We re-write Eq.~(\ref{eq:log_linear_fit}) in terms of the total surface density, $\Sigma$, and the gas fraction, $f_{\rm g}$, and obtain
\begin{equation}
\beta (\Sigma , f_{\rm g}) = \beta_0  \left(\frac{\Sigma}{1 \, M_\odot \, {\rm pc}^{-2}} \right)^{-\mu} f_{\rm g}^{\nu - \mu}\,,
\end{equation}
giving a dependence on the gas fraction, $\propto f_{\rm g}^{-0.99}$. The fraction of the wind launched as function of radius is given by,
\begin{eqnarray}
\frac{{\rm d} f_{\rm w}}{{\rm d} (r/R_{\rm d})} &=& \frac{2 \pi R_{\rm d} \beta(r) \dot{\Sigma}_\star(r)}{\dot{M}_{\rm wind}} \nonumber \\
&=& (n-\mu)^2  \left( \frac{r}{R_{\rm d}} \right) \exp\left[ -(n-\mu) r/R_{\rm d} \right] \, , \label{eq:frac_wind}
\end{eqnarray}
which gives the differential rate of production of the star-formation driven wind, normalised by the total wind. This function is plotted in Fig.~\ref{fig:disc_wind}. At large radii the star formation is most effective at driving a wind, but the net contribution to the galaxy outflow is limited by the low rate of star formation there. Conversely at small radii the wind is limited by the small area of the disk, and so it is at intermediate radii where the local mass loading equals that of the galaxy as a whole.

We can characterise this further by defining a wind radius $R_{\rm w}$ by
\begin{equation}
\hat{\beta} = \beta \left( \Sigma(R_{\rm w}), f_{\rm g} \right)\,,
\end{equation}
that is, $R_{\rm w}$ is that radius in the galaxy where the local mass loading, $\beta=\dot\Sigma_{\rm ej}/\dot\Sigma_\star$, equals the total mass loading of the entire galaxy,
$\hat\beta=\dot M_{\rm wind}/\dot M_\star$. The wind radius for the galaxy is then given by,
\begin{eqnarray} \label{eq:r_wind}
R_{\rm w} &=& \frac{2}{\mu}\log \left(\frac{n}{n-\mu}\right) R_{\rm d} \\
&\approx& 3.0 R_{\rm d} \, ,
\end{eqnarray}
where we have substituted in $n=1.4$ for the exponent in the KS relation, and used the values for $\mu$ from Eq.~(\ref{eq:mu_fit}). For the Milky Way, a disk scale length of $R_{\rm d} = 2.5 \, \rm kpc$ gives a wind radius of $R_{\rm w} = 7.5 \, \rm kpc$, inside the solar radius but outside the galactic bulge. We have neglected the fact that there will not be any star formation far out in the disk if the gas surface density drops too low, as well as the presence of a bulge, where there may be little gas and hence also little star formation. This will lead us to overestimate the wind in the tails of Fig.~\ref{fig:disc_wind}.

To parameterise feedback in terms of the circular velocity, $V_{200}$, we apply Eq.(\ref{eq:MMW_sigma}) and use our fiducial values of $\beta_0, \mu, \nu$ and $f_{\rm g}$, to find
\begin{eqnarray}
\hat{\beta} &=& \beta_0 \left( \frac{n}{n-\mu} \right)^2 \left(\frac{\Sigma_0}{1 \, M_\odot \, {\rm pc}^{-2}} \right)^{-\mu} f_{\rm g}^{\nu - \mu} \label{eq:beta_Sig0} \\
&\approx& 10  \left( \frac{\beta_0}{13} \right) \left({f_{\rm g}\over 0.2}\right)^{\nu - \mu} \left( \frac{j_{\rm d}}{m_{\rm d} v_{\rm d}} \right)^{2\mu} \times \nonumber \\
&& \left[ \left( \frac{\lambda}{0.05} \right)^{-2} \left(\frac{V_{200}}{155\, \rm km \, s^{-1}}\right)  \left(\frac{m_{\rm d}}{0.03}\right) \frac{H(z)}{H_0} \right]^{-\mu} \, , 
\label{eq:hat}
\end{eqnarray}
where we also assumed $H_0 = 71 \rm \; km \, s^{-1} \, Mpc^{-1}$ \citep{Freedman_2001}.

To convert to disk properties we can eliminate the spin parameter with
\begin{equation}\label{eq:Rd_angmom}
R_{\rm d} = \frac{\lambda V_{200} }{\sqrt{200} H(z)} \left( \frac{j_{\rm d}}{m_{\rm d} v_{\rm d}} \right) \, .
\end{equation}
Setting $j_{\rm d}/m_{\rm d}$ and $v_{\rm d}$ as unity, however, yields a MW with a rather low circular velocity ($155$~km s$^{-1}$) and scale length considerably higher.

The formation of the baryonic disk can increase the rotation velocities from $V_{200}$ both directly and indirectly. The baryons make their own contribution to the gravitational potential, and can also induce changes in the profile of the dark matter, for example due to adiabatic contraction (e.g. \citealp{Mo_2010}). Even without baryons, there will be some adjustment to $v_{\rm d}$ due to the non-isothermal nature of halos \citep{NFW_1997}, i.e. a dependence on the concentration parameter. Here we will take $v_{\rm d} = 1.29$ to give a circular speed of $V_{\rm d} = v_{\rm d} V_{200} = 200 \; \rm km \, s^{-1}$, similar to the value of the MW (\citealp{Dehnen_1998, Flynn_2006}, but see also \citealp{Reid_2009} that has the speed closer to $250 \; \rm km \,  s^{-1}$). 

Having set the circular speed, the disk scale length is implied by the specific angular momentum fraction in Eq.~(\ref{eq:Rd_angmom}). For the $2.5$ kpc disk of \cite{Flynn_2006} we set $j_{\rm d}/m_{\rm d}=0.42$, i.e. the disk is preferentially formed of the low angular momentum baryons. A  possible reason for the lower specific angular momentum is the delayed collapse of baryons in the disk due to photo-heating (since disks grow in an inside-out manner, with the low angular momentum material accreted first), \cite{Navarro_1997}. 

Finally we should mention that the spin parameter of the MW may differ from $0.05$, and indeed recent simulations that remove transient objects from halos have suggested halos have a smaller $\lambda$ (e.g. \citealp{Bett_2007}), however we have made no account for this as it is outside the scope of this model.

With these new parameters, the MW disk has a more realistic higher surface density, and Eq.~(\ref{eq:hat}) becomes
\begin{eqnarray}
\hat{\beta} &\approx& 0.31  \left( \frac{\beta_0}{13} \right) \left({f_{\rm g}\over 0.2}\right)^{\nu - \mu} \times \nonumber \\
&& \left[ \left(\frac{V_{\rm d}}{200\, \rm km \, s^{-1}}\right)^3 \left( \frac{R_{\rm d}}{2.5 \; \rm kpc} \right)^{-2}  \left(\frac{m_{\rm d}}{0.03}\right) \frac{H_0}{H(z)} \right]^{-\mu}\,. \label{eq:beta_vcirc}
\end{eqnarray}

The normalisation and scaling with $V_{\rm d}$ we find are somewhat below our expectations for supernova feedback. For a Milky-Way like halo, the star formation would remove less than one solar mass of gas for every solar mass of stars formed ($\hat\beta\sim 0.31$). Nevertheless, halos with smaller circular velocities with the same disk radius and disk mass fraction show increasingly effective feedback, $\hat\beta\propto V_{\rm d}^{-3.4}$, a similar scaling to energy conserving winds (e.g. \citealp{Stringer_2011}). Note that the power-law dependence on $V_{\rm d}$ is somewhat stronger than the value of -1 found by \citet{Hopkins_2011}. Those authors also found an exponent of $-0.5$ for the dependence of mass loading on surface density, which is weaker than our exponent in Eq. (\ref{eq:beta_vcirc}) of $\hat\beta\propto \Sigma^{-\muEst }$. Whilst the agreement between these simulations is not particularly good, this is perhaps not surprising given that they are performed with some different physics, at different resolutions and using different hydrodynamical schemes. 

Despite the appeal of the above framework in supplying us with predictions for the mass loading in terms of redshift and the disk properties, there is a caveat here in our adjustment of $j_{\rm d} / m_{\rm d}$ and $v_{\rm d}$ to match the observed MW. Although we can derive this from observations for the MW, and the mechanism for this appears to be understood, it would be erroneous to suggest we have a consistent model for this, and current numerical simulations such as those of \cite{Scannapieco_2011} have yet to converge on the properties of a disk for a single halo. Most concerning is that these quantities almost certainly have some implicit dependence on halo mass and thus there should be a corresponding adjustment to the scaling relation in Eq.~(\ref{eq:beta_vcirc}).

\subsection{Dependence from observed data}\label{sec:beta_TF}

Given the approximate ingredients required to construct the formalism of the previous section, it is interesting to ask whether we can parameterise our fit to the mass-loading, Eq. (\ref{eq:beta_Sig0}), with purely observational estimates, i.e. to compute the disk surface density with from observed disk properties, side-stepping the models of \citet{Mo_Mao_and_White_98}.

One particularly attractive method is to invert Eq.~(\ref{eq:md}) to write the surface density in terms of the disk radius $R_{\rm d}$ and mass $M_{\rm d}$, where the latter can be estimated from the circular velocity of the disk with the Tully-Fisher relation \citep{Tully_Fisher_1977}. A recent calibration of the baryonic Tully-Fisher relations gives $M_{\rm d} = 8 \times 10^{10}\, M_\odot (V_{\rm max}/200~{\rm km}~{\rm s}^{-1})^4$ \citep{Trachternach_2009}, application of which gives
\begin{eqnarray}
\hat{\beta}_{\rm TF} &=& 0.31 \left( \frac{\beta_0}{13} \right) \left({f_{\rm g}\over 0.2}\right)^{\nu - \mu} \times \nonumber \\
&& \left[  \left(\frac{V_{\rm d}}{200\, \rm km \, s^{-1}}\right)^4  \left(\frac{R_{\rm d}}{2.5 \; \rm kpc}\right)^{-2} \right]^{-\mu} \, , \label{eq:beta_TF}
\end{eqnarray}
which is very close to the relation in Eq.~(\ref{eq:beta_vcirc}), including normalisation and the $R_{\rm d}$ scaling. The difference is in the exponent of $V_{\rm d}$, and the dependence of Eq.~(\ref{eq:beta_vcirc}) on  $m_{\rm d}$, which implicitly depends upon $V_{\rm d}$ as well.

In principle it is possible to calculate the mass fraction in the disk from the stellar mass to halo mass function using an abundance matching approach, 
which would relate $m_{\rm d}$ to $V_{200}$. A single power law, $m_{\rm d} \propto M \propto V_{200}^{3}$ is a good fit, although from Eq. (\ref{eq:betascale}) we see there is a dependence on the faint end slope of the stellar mass function (and at higher masses a broken power law may be more appropriate, e.g. \citealp{Yang_2003, Moster_2010, Guo_2010}). Substituting this relation for $m_{\rm d}$ in Eq.~(\ref{eq:beta_TF}) then yields $\hat\beta\propto V_{200}^{-6.9} R_{\rm d}^{2.3}$ 
 versus $\hat{\beta}_{\rm TF} \propto V_{\rm d}^{-4.6} R_{\rm d}^{2.3}$  from Eq.~(\ref{eq:beta_TF}) (taking $\mu=\muEst$ for both). Finally we can try to eliminate the dependence on $R_d$, assuming  $R_{\rm d} \propto M_{\rm d}^{0.15}$, as inferred by \cite{Shen_2003}. This yields a scaling of $\hat\beta\propto V_{\rm d}^{-4.8}$ versus $\hat{\beta}_{\rm TF} \propto V_{\rm d}^{-2.5}$. The difference between these scalings is due to the discrepancies between the modelled and observed slope for the Tully-Fisher relation and the uncertainty in modelling the disk mass fraction.

Although both our scalings are strongly dependent on $V_{\rm d}$, our $\beta$ values were all in the range 0.01-4, so the change in feedback acts more like a switch. At low disk circular velocities $V_{\rm d} \ltsima 140\; \rm km \, s^{-1}$ the feedback is high ($1<\beta < 4$) to and at the higher disk velocities the feedback shuts off, all over a relatively small range in $V_{\rm d}$.

To summarise, we have developed two approaches to analyse the mass loading for a galaxy based upon our estimates for the mass loading in our ISM patches. In Section \ref{sec:beta_MMW} we take an analytic approximation to the properties of disk in their host halos which allows us to trace the feedback with redshift. This does, however, require us to make assumptions about the scaling of the gravitational contribution of the baryonic disks and the preferential accretion of low angular momentum baryons, neither of which are fully understood. Section \ref{sec:beta_TF} has bypassed these model concerns by parameterising the galaxies using the observed disk mass-velocity relation to directly apply the mass loadings. One price for this is the loss of the dependence on redshift and the cosmological evolution.

Although these two approaches lead to different scalings, they do give a consistent normalisation for the feedback in the MW at redshift zero. In principle, one way to test this formalism is to apply it in phenomenological models such as \galform , where such parameters as $j_{\rm d}, v_{\rm d}$ and $m_{\rm d}$ are followed. We discuss this comparison further in the next section.

\subsection{Comparison to cosmological models}

We are now in a position to compare the outflow rate we measured in our high resolution simulations with values assumed in semi-analytic models such as \galform\ \citep{Cole_2000}. The feedback prescription for the original \galform\ was
\begin{equation}
\beta = \left({V_{\rm d}\over V_{\rm hot}}\right)^{-\alpha_{\rm hot}} \, ,
\end{equation}
with values in the reference model of $V_{\rm hot} = 200\; \rm km \, s^{-1}$ and $\alpha_{\rm hot}=2.0$. These models give a slope to the faint end of the galaxy luminosity function, $\alpha\approx -1.5$. More recent models such as \citet{Bower_2006} have used $\alpha_{\rm hot}=3.2$ for a good match to the $b_{\rm J}$ and K-band galaxy luminosity functions. These can be compared with our exponents from the previous paragraph, $\alpha_{\rm hot}=4.8$ and $\alpha_{\rm hot,TF}=2.5$, which bracket the value used by \cite{Bower_2006}. For the normalisation, \cite{Cole_2000} parameters yield $\beta_{200}=1.0$ ($\beta$ for a disk of $V_{\rm d}=200$~km~s$^{-1}$), whilst the \cite{Bower_2006} parameters give $\beta_{200} \approx 17$ (although this drops to 12 using updated cosmological parameters, see \citealp{Bower_2011}) as compared with our value of $\hat\beta_{200}=0.31$. The net mass loading for MW like galaxies obtained from our simulations is less than that assumed by \citep{Cole_2000} by about a factor 2, and considerably less than assumed by \cite{Bower_2006}. 

It is also interesting to consider whether the values of $\beta$ should rise in starburst galaxies, where the star formation rate may be significantly above the normalisation of the Kennicutt-Schmidt relation. Although our higher star formation rate simulations do show higher values of $\beta$ (see Appendix A), this is only by a factor of 2, with $\beta$ still falling at high gas surface densities. This suggests that the mechanism for galaxies to stay at high mass loadings is to remain in a state with relatively low surface densities (e.g. \citealp{Read_2006}).

An alternative formulation of feedback in semi-analytics, suggested by \citet{Bower_2011}, is to attempt to match only the observable portion of the stellar mass function rather than trying to match a slope that goes to arbitrarily faint galaxies. For example, a model with a constant wind speed (from the disk) ultimately produces a faint end slope that is identical to that of the halo mass function. In an intermediate mass range, however, the effects of the gravitational potential causes material to be recycled back into the galaxy, producing a characteristic flat portion to the galaxy stellar mass function. By tuning the value of the wind speed, a nearly flat stellar mass function can be achieved over a restricted range. Although this mechanism cannot be extended to arbitrarily faint galaxies (which may be suppressed by other mechanism for example by re-ionization), it does provide a good fit to the observations with a constant $\beta \approx 8$ over this portion of the mass function. 

In contrast to some of the predictions of semi-analytic models are the smaller estimates for the normalisation for mass loading found by hydrodynamic simulations. \cite{Oppenheimer_2010} use a $\beta = 2$ and a $v_{\rm wind}=680 \; \rm km \, s^{-1}$ to recreate the $z=0$ mass function. These simulations are at low resolution with the wind particles partially decoupled from the surrounding gas, making them more comparable to semi-analytic models. Fully hydrodynamical simulations where the wind is coupled to the surrounding ISM are much harder to interpret. Resolution of these issues is beyond the scope of this paper, but better understanding of the differences between semi-analytic models and hydro simulations is clearly required.

In terms of the observed MW, \cite{Wakker_2008} estimates the mass accretion rate to be $0.4\; \rm M_\odot \, yr^{-1}$ from infalling high velocity clouds. If this is to be combined in a steady state model of a MW with non-negligible star formation, then $\beta_{200} \lesssim 0.4$, so there is some tension between the observed star formation of the MW and the semi-analytic models that would reduce its baryon fraction, and our simulations lie nearer the low observed estimates.

One option is that the semi-analytic models consistently over-estimate the $\beta_{200}$ required. In particular, there are significant degeneracies between $\beta_{200}$ and the exponent $\alpha_{\rm hot}$. Moreover, many models assume that the wind scaling has a fixed energy efficiency ($\eta_T$) and do not correctly account for the recapture of gas ejected from low mass galaxies (see \citealp{Bower_2011} for further discussion). It is entirely plausible that a careful search of parameter space may revel strongly mass dependent solutions much closer to those found here.

On the hydrodynamical side, there are a number of physical processes that we neglected that may nevertheless be important. In terms of the gas phases we have included, the inhomogeneous metallicity will make an adjustment to the cooling, and larger scale effects such as a full 3-dimensional galactic potential along with shear and features such as bars and spiral arms will also play a role in shaping the ISM. However, it is not apparent why either of these effects will change the overall mass leaving the disk. In terms of the stellar populations we could explore the star formation distribution in terms of the correlation with molecular clouds and also the clustering of stars, which may allow the explosions to strip more material, but this is unlikely because SNe are delayed sufficiently to diffuse out of their parent clouds. The large scale radiation field may provide an additional mechanism to accelerate the wind \citep{Murray_2005, Hopkins_2011}, however in our simulations the thermal energy of the hot material in the disk already provides sufficient velocities to escape the disk.

Potentially the largest discrepancy we have identified is the inconsistency of the distribution of SNe with the gas evolution, i.e. matching the scale height of star formation with the new scale height of the disk. It may even be possible to make the simulations completely self consistent by matching the star formation rate to the turbulent structure of the ISM, in a manner such as that envisaged by \cite{Krumholz_2005}.

Future simulations could also include the cold phase of the ISM by including radiative cooling below $10^4\; \rm K$. On its own this would tend to reduce $\beta$, since a cold phase removes material from the warm phase it would not directly increase the mass loading, however the physics of this brings in other processes such as self gravity, magnetic fields, and cosmic rays (which may be dominant at these scales). Magnetic fields in particular seem a candidate for entraining more material into the wind, although simulations such as \cite{Hill_2012} do not find it to play a significant role.

Overall, whilst we will include the above physical processes in future work, we suspect that these processes will not radically alter the mass-loading or significantly change the scalings we have found.

\section{Conclusions}
\label{sect:conclusions}

In this paper we have constructed numerically well converged simulations of a simplified two-phase interstellar medium model, in which an initially isothermal and hydrostatic disk gets disrupted and heated by individual supernovae. By not simulating the cold phase of the ISM we avoided the need to introduce significantly more physical ingredients which require heavy algorithmic approximations and/or fragile recipes. By restricting our simulation volume to only a small section of a disk, we achieve sub-parsec resolution, and are able to investigate the dependence of the outflow on the parameters of the disk. We have included fixed gravity corresponding to our hydrostatic initial conditions, star formation that follows the Kennicutt-Schmidt relation, hydrodynamics and a cosmological cooling function. On scales outside the volume, the host disk galaxy for this toy model is reduced to the parameters of gas surface density, gas fraction and star formation efficiency normalised by the Kennicutt-Schmidt relation. 

Our simulations demonstrate the ability of supernovae to launch a galactic wind vertically from a disk, although we do not follow the subsequent evolution of the material in the halo. The supernovae create a turbulent ISM with very distinct hot and warm phases, due to the strong transition of the cooling function at $10^4\; \rm K$. These phases exist in order-of-magnitude pressure equilibrium, with the warm material squeezed into dense lumps, and the excess thermal energy of the hot material causing it to accelerate away from the disk. In section \ref{sec:rarefaction} we compare this to a rarefaction process, with the hot ISM escaping to an IGM which is comparatively sparse and pressure-free. Such a model naturally leads an outflow with speed increasing with height above the disk but density decreasing.

The hot outflow entrains colder ISM gas from the disk, that may have relatively high metallicity.  The hot gas rushes past this cloudy medium producing characteristic tails. Such interfaces may be the cites where lower ionisation lines are produced. In section \ref{sec:MockAbsorb} we explore this further by calculating the normalised cross section of different temperature phases in our simulations, where we see the velocity distribution of the cooler gas is significantly beneath that of the escaping material.

In a given snapshot the precise features of our simulations vary greatly due to turbulence and the stochastic nature of supernovae, therefore we examine several global properties which are less sensitive, such as the disk pressure, cooling rate as a fraction of the mean energy injection rate, disk scale height and mass ejection. These reveal a disk that rapidly evolves to higher porosity before reaching a state with an approximately constant mass ejection rate. This evolution of porosity is broadly reminiscent of the model by \citet{Silk_2001}. 

We perform a range of simulation to investigate the dependence of the mass loading on gas surface density, gas fraction, and star formation efficiency, and fit the resulting trends with power laws. Our mass loadings lie in the range $0.01$-$4$, suggesting a switch from a low to a high feedback regime at $V_{\rm d} \approx 140 \; \rm km \, s^{-1}$. We find little dependence on the normalisation of the star formation relation but a significant dependence on the gas fraction and surface density. The latter two can be combined to explain the bulk of the trends as depending on the total surface density of the disk. At high surface densities we find low mass loading and a high effective wind speed. At low surface densities the reverse is true, and there is an additional contribution due to an increase of the fraction of energy radiated by cooling gas. In Section \ref{sec:CharacTemp} we present a simple model where SNe blasts stall as they run into clouds swept-up by previous explosions that are so dense that they cool very efficiently predicts that mass loading depends on gas surface density and gas fraction as $\beta=\dot\Sigma_{\rm wind}/\dot\Sigma_\star \propto \Sigma_{\rm g}^{-8/11}\,f_{\rm g}^{4/11}$. These scalings are very close to those we find from simulations with high star formation rate, $\beta\propto \Sigma_{\rm g}^{-0.82}\,f_{\rm g}^{0.48}$ and weaker (in terms of surface density) than that for the pure Kennicutt relation, $\beta\propto \Sigma_{\rm g}^{-\muEst}\,f_{\rm g}^{\nuEst}$. Our prediction for the mass loading in the solar neighbourhood is that each supernova results in an ejection of around $50 \; \rm M_\odot$ of gas, or a $\beta \sim 0.5$, slightly above $0.3$, our the average for the MW as a whole.

The relationship between the wind velocity and thermalisation efficiency exhibits a more complex relationship to the disk properties than that of the mass loading. The thermalisation efficiency appears to show a dependency on both the surface density and the gas fraction, and correspondingly the wind velocity does not show a straightforward power law implied from a constant efficiency model. For high surface densities and low gas fractions, an approximate $40\%$ of the injected energy is converted into the outflow's thermal, turbulent and kinetic energy components, although we will underestimate the cooling outside our simulation volume.

We employ the scaling relation obtained from the simulations to calculate the net mass loading, $\hat\beta=\dot M_{\rm wind}/\dot M_\star$, of an exponential disk galaxy with constant gas fraction. Using the \cite{Mo_Mao_and_White_98} scaling relation between disk and halo, we obtain a scaling with circular velocity of $\hat\beta\propto V_{\rm d}^{-4.8}$, stronger than either energy or momentum-driven winds. Using the observed Tully-Fisher relation we find a weaker dependence, $\hat\beta\propto V_{\rm d}^{-2.5}$. This compares well with recent semi-analytic models which assume $\alpha_{\rm hot} \in [2.0, 3.2]$. 

The normalisation of our net mass loading at redshift $z=0$ for a Milky-Way like galaxy is significantly lower than assumed in recent phenomenological models, although these models appear to have some degeneracy between the exponent and the normalisation, which we will exploit in future work. Notably the mass loading only increases weakly with star formation rate but decreases strongly with surface density, so for starburst galaxies the feedback may be less efficient. Interestingly, our estimated normalisation is comparable with inferred values of outflow for the MW based upon the observed accretion and star formation. If indeed there is a higher mass loading, it will require supernovae to heat a larger mass of material to a lower temperature, or for the hot outflow to entrain a larger fraction of the warm ISM gas.

The scaling we find sets the investigation of galaxy winds on a new footing, providing a physically motivated sub-grid description of winds that can be implemented in cosmological simulations and semi-analytic models.

\section*{Acknowledgements}   
Peter Creasey would like to acknowledge the support of an STFC studentship. The authors would like to thank Martin Stringer, Tom Abel, Crystal Martin, Claudia Lagos and Andrew Pontzen for helpful discussions. We would also like to thank the anonymous referee for comments which substantially improved this paper. The calculations for this paper were performed on the ICC Cosmology Machine, which is part of the DiRAC Facility jointly funded by STFC, the Large Facilities Capital Fund of BIS, and Durham University. The \flash\ software used in this work was in part developed by the DOE-supported ASC/Alliance Center for Astrophysical Thermonuclear Flashes at the University of Chicago. This research was supported in part by the National Science Foundation under Grant NSF PHY11-25915.

\bibliographystyle{mn2e}
\bibliography{bibliography} 
\bsp
\begin{appendix}
\section{Convergence and Parameter fits}\label{sec:convergence}

In this appendix we investigate the convergence properties of our simulations, along with the dependence upon some of the numerical and physical parameters such as the box size, simulation end time, the star formation rate, the cooling function and the energy of each supernovae. We also include evolution graphs showing the fits to to the mass loading which are central to this work.

We begin by describing a set of simulations where we run an alternative star formation rate, which is compared to the main set of simulations in Section \ref{sec:fitparams}. In this parameterisation, the surface density of star formation is 
\begin{equation}\label{eq:KStdyn}
\dot{\Sigma}_\star = 0.1 \Sigma_{\rm g} / t_{\rm dyn}\, ,
\end{equation}
(more commonly used in cosmological simulations), which is appropriate for a marginally Toomre stable disk \citep{Toomre_1964}, i.e. the vertical dynamical time is close to the orbital time scale. Such prescriptions are discussed thoroughly in \citet{Schaye_and_Dalla_Vecchia_08}, where they show that with self-regulating feedback this will recover an approximate Kennicutt-Schmidt relation of $\Sigma_{\rm g}^{3/2}$.

If we apply Eq.(\ref{eq:KStdyn}) to the warm disk of our initial conditions, however, we generally have a much higher star formation rate due to the short dynamical time, which is equivalent to saying the HI disk is not Toomre stable. If we substitute in the $t_{\rm dyn}$ from Eq.~(\ref{eq:tdyn}) we have a star formation rate of 
\begin{eqnarray}
\dot{\Sigma}_\star &=& 2.6 \times 10^{-3} \left(\frac{f_{\rm g}}{0.1} \right)^{-1} \left( \frac{\Sigma_{\rm g}}{10 \; \rm M_\odot \, pc^{-2}}\right)^{\frac{1}{2}}  \times \nonumber \\
& & \Sigma_{\rm g1}^{1.5} \; \rm M_{\odot} \, kpc^{-2}\, yr^{-1} \, ,
\end{eqnarray}
which we can see from the leading coefficient is an order of magnitude higher than Eq. (\ref{eq:KS}), alhough there is some residual dependence on $f_{\rm g}$ and $\Sigma_{\rm g}$.To some extent, however, this simulates the conditions more relevant to a starburst galaxy. 

\begin{figure}
\centering
\includegraphics[width=\columnwidth]{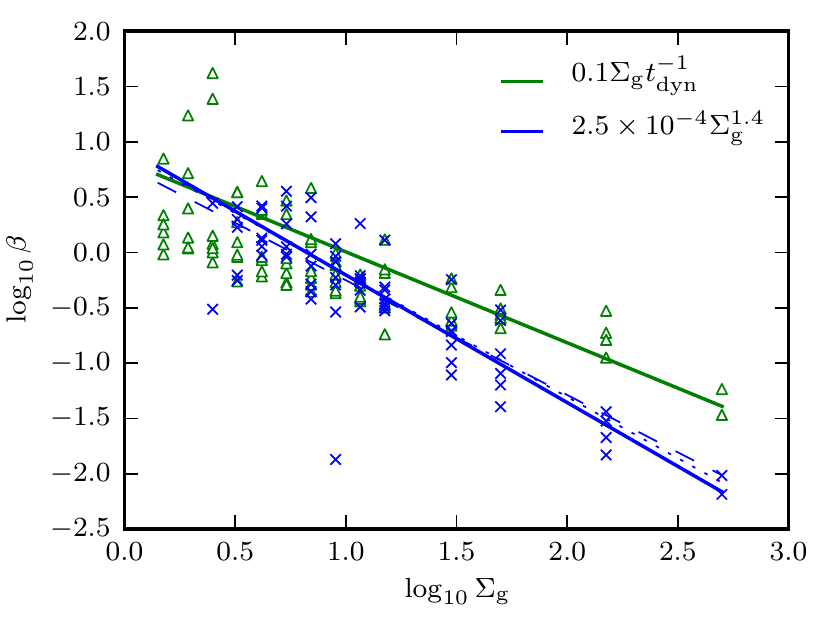}
\caption{Mass loading $\beta$ as a function of $\Sigma_{\rm g}$ for the star formation laws in Eq. (\ref{eq:KS}) (\emph{blue crosses}) and Eq. (\ref{eq:KStdyn}) (\emph{green triangles}). \emph{Solid lines} indicate the best fit for $f_{\rm g}=0.1$ (see text for exact parameterisation). \emph{Dashed} and \emph{dotted blue} lines show the fit when varying the end time of the data used for the fit by $\pm 3$ Myr.}
\label{fig:SFRdep}
\end{figure}

In Fig.~\ref{fig:SFRdep} we investigate the effect of altering the star formation law from Eq.(\ref{eq:KS}) to Eq.(\ref{eq:KStdyn}), where the latter in general has much higher star formation rates due to the short vertical dynamical time of the disk. At low gas surface densities more simulations were possible due to the higher star formation rates. The best fit to the former was given in the main text, whilst the best fit to the latter is
\begin{equation}
\beta \sim  (20 \pm 8) \Sigma_{\rm g1}^{-0.82 \pm 0.07} f_{\rm g}^{0.48 \pm  0.08} \, .
\end{equation}
The effect of increasing the star formation rate flattens the dependency on $\Sigma_{\rm g}$ and increases the dependency on $f_{\rm g}$, very close to the values predicted in Eq.~(\ref{eq:beta_surf}), which is to be expected as gravity is much less important in these simulations (see also the discussion in Section \ref{sec:fitparams}). The relative insensitivity of $\beta$ to the order of magnitude change in $\dot{\Sigma}_\star$ can be explained by the fact that the outflows are normalised by the star formation rate, so although those simulations have much higher outflows, the outflow per supernova deviates by a much smaller amount. The higher star formation rate runs can also be seen to have less scatter, as they are less susceptable to the Poisson noise of individual SN events.

\begin{figure}
\centering
\includegraphics[width=\columnwidth]{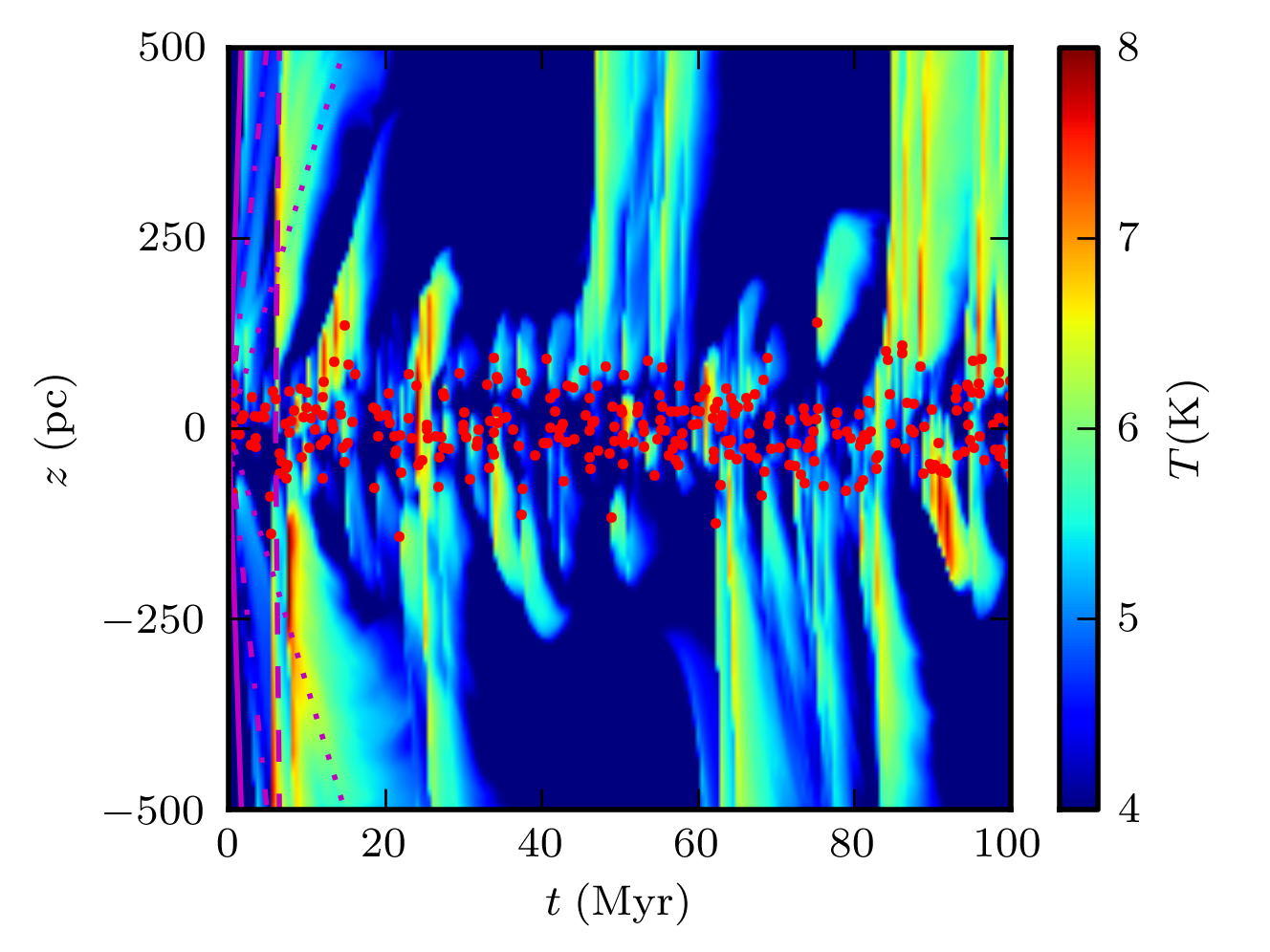}
\includegraphics[width=\columnwidth]{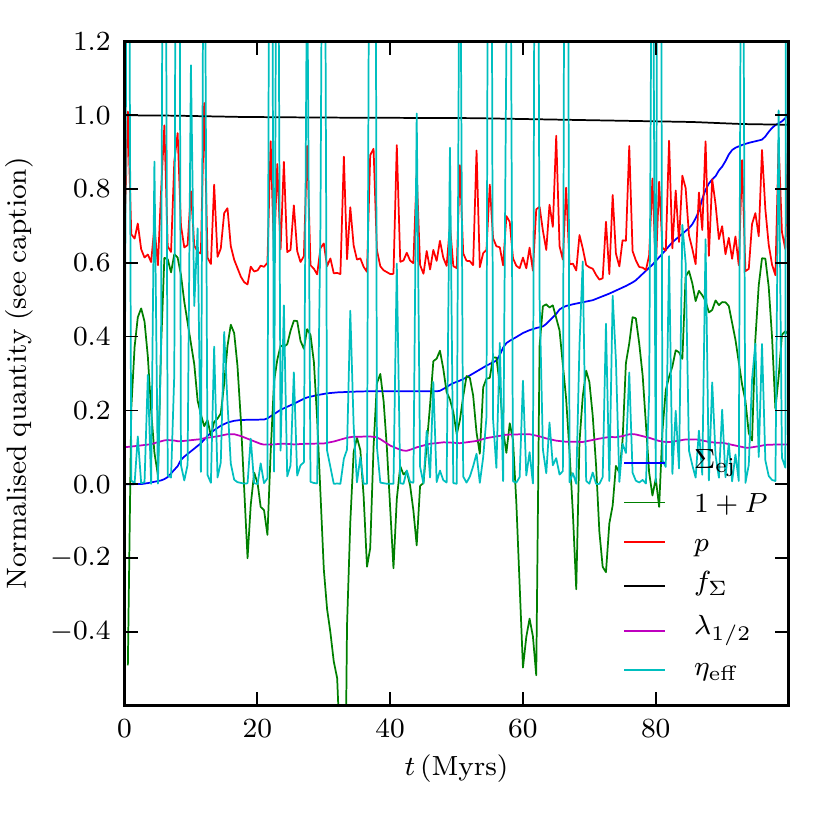}
\caption{As for figures \ref{fig:fiducial_convergence} (\emph{lower panel}) and \ref{fig:height_time} (\emph{upper panel}) but for a run of 100 Myr. In the lower panel the surface density ejected (\emph{dark blue line}) has been scaled to $0.3 \; \rm M_\odot \, pc^{-2}$.}
\label{fig:long_time}
\end{figure}

Fig. \ref{fig:SFRdep} also illustrates the effect of the simulation end time on our estimate for the gas surface density dependency, by varying by $\pm 3$ Myr the final snapshot which is used to construct the fits for $\beta$ (for the normal star formation rates). This shows little effect, a result of the outflow rates being (on average) very close to linear in these simulations. We perform a corresponding fit for the fiducial parameters in Fig. \ref{fig:long_time} where we simulate 100 Myr to check that the outflows we see are not a transient phenomena and continue after the 20 Myr of our simulations. The box width in this simulation was $200$ pc, so we expect to see more stochasticity, and indeed we see fluctuations lasting many Myrs, such that the outflow estimated from a single 20 Myr window could show a deviation of a factor of a few. This is probably the main reason for the scatter in Fig. \ref{fig:betafits}.

\begin{figure}
\centering
\includegraphics[width=\columnwidth]{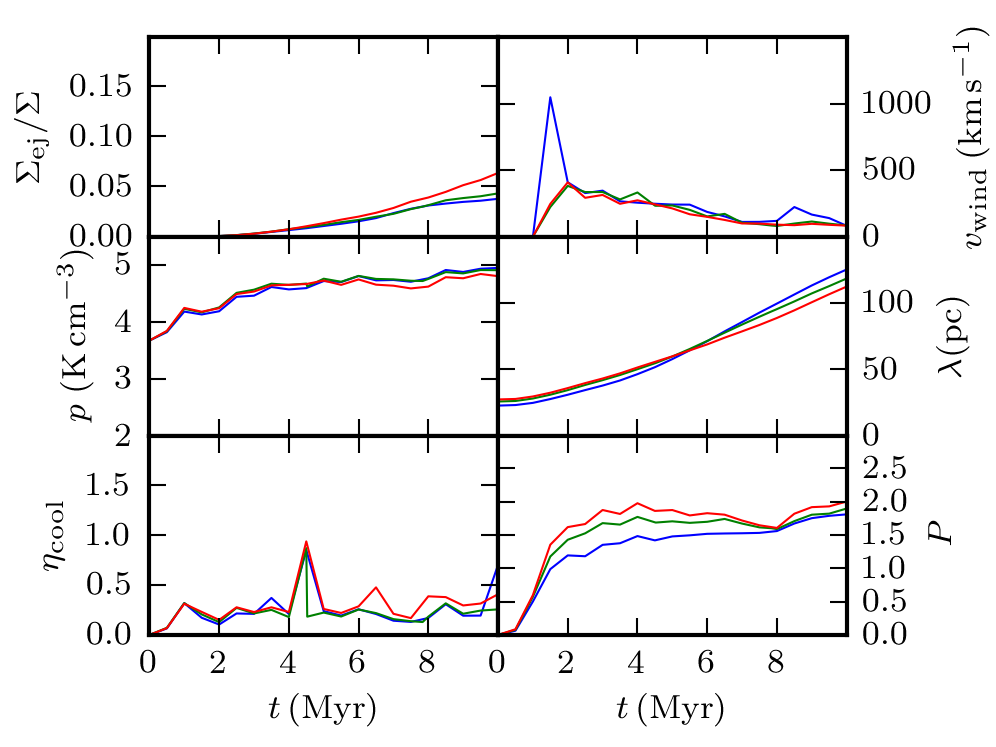}
\caption{Numerical convergence of a high star formation rate run at resolution of L2, L3, L4 (cell size of $6.3$, $3.1$ and $1.6$~pc and shown in red, green and blue respectively).
\emph{Upper left panel} shows the fraction of gas that has left the simulation volume, \emph{middle left} indicates the mean pressure in the simulation volume. \emph{Lower left panel} shows the rate of cooling as a fraction of the mean supernovae energy injection rate, \emph{upper right} shows the mean wind velocity, \emph{middle right} shows the scale height of the disk and \emph{lower right} shows the evolution of the porosity in the simulation. The red and green curves follow each-other closely, indicating good convergence.}
\label{fig:fiducial_convergence}
\end{figure}

We also test how well our simulations are converged w.r.t. the resolution by taking one of the high star formation rate runs and re-simulating it at the three resolutions L2, L3 and L4 (corresponding to a cell size of $6.3$, $3.1$ and $1.6$~pc, respectively). In Fig.~\ref{fig:fiducial_convergence} we show six different parameters, the fraction of surface density ejected, the mean pressure in the simulation volume, rate of cooling (as a fraction of the supernova energy injection), the porosity and scale heights of the disk and the wind velocity for three different resolutions. All of these properties appear to be well converged with respect to resolution, with the possible exception of the porosity and the disk scale height at the lowest resolution. With respect to the scale height it is notable that there is some error even at the initial snapshot due to the coarseness of the grid in this case. 

\begin{figure}
\centering
\includegraphics[width=\columnwidth]{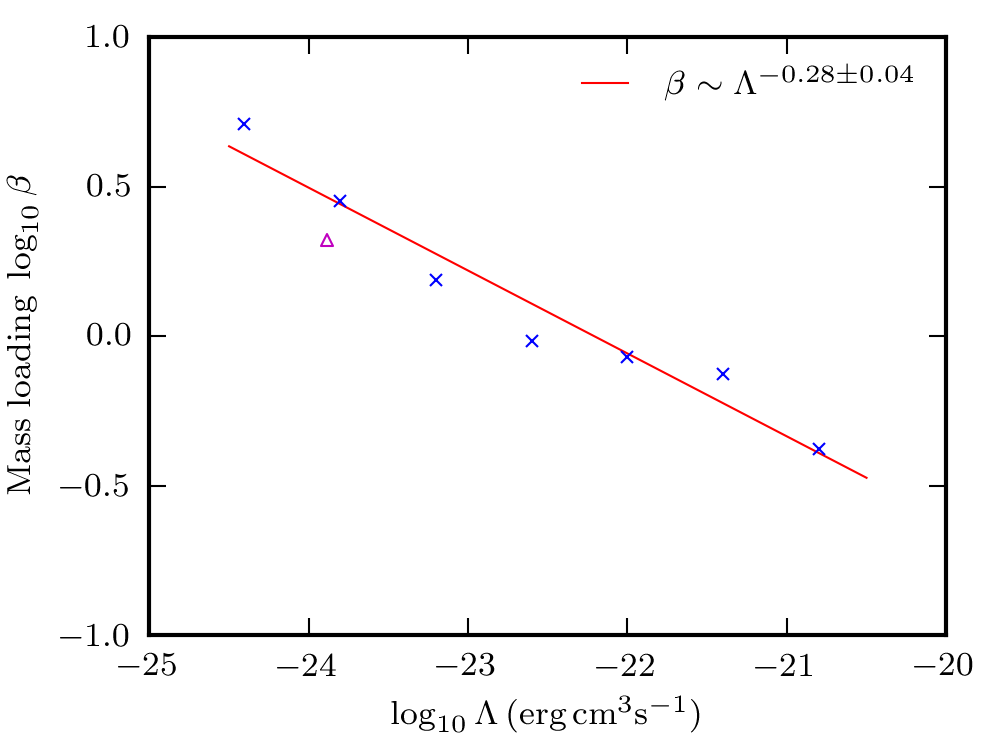}
\caption{Dependence of the mass loading parameter $\beta$ on the cooling rate. \emph{Blue crosses} show the estimated $\beta$  for different values of $\Lambda$ for
Heaviside-shaped cooling function ($\Lambda=10^{-22} \, \rm erg \, cm^3 \, s^{-1}$ is the fiducial value). For comparison, the \emph{maroon plus} indicates the value of $\beta$ calculated with the \citet{Sutherland_1993} cooling function, at the minimum value of this function (Eq. (\ref{eq:SD_min})). } 
\label{fig:cooling_convergence}
\end{figure}

\begin{figure}
\centering
\includegraphics[width=\columnwidth]{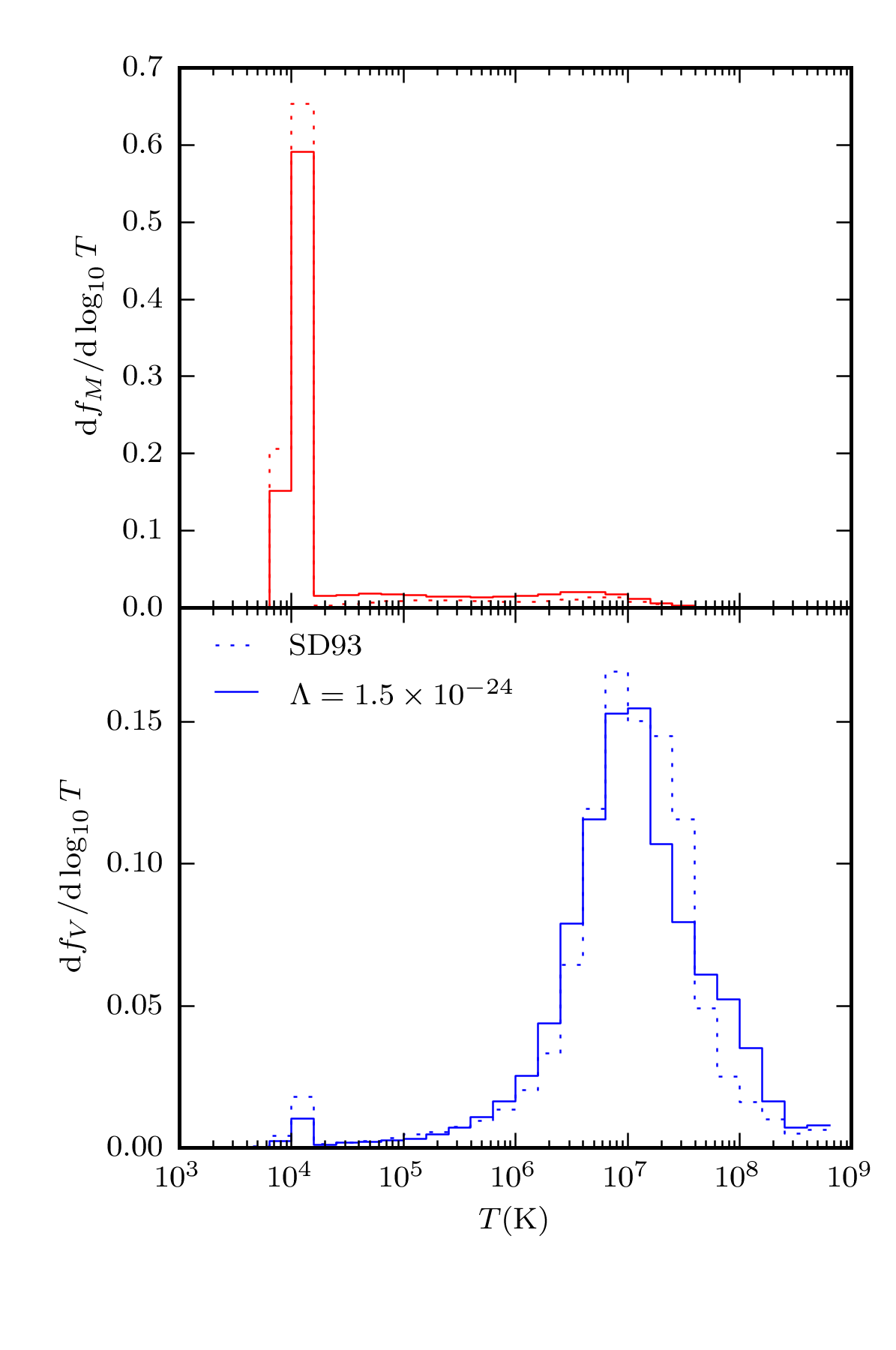}
\caption{\emph{Upper panel}, mass fraction of the gas in different temperature phases, \emph{solid, dashed lines} refer to the $\Lambda=1.5 \times 10^{-24}$~erg~cm$^3$~s$^{-1}$ Heaviside cooling function,  and the \citet{Sutherland_1993} cooling functions respectively. \emph{Lower panel} shows the corresponding volume fractions. The fraction in each phase appears very similar, with the SD93 cooling function showing a slightly narrower temperature distribution of the hot phase by volume.}
\label{fig:sd_comparison}
\end{figure}

We explore the importance of cooling, both in broad terms about the dependence on the magnitude of the cooling, and also upon our specific choice of cooling function. In Fig.~\ref{fig:cooling_convergence} we look at the dependency of $\beta$ for the previous simulation on the magnitude of the cooling function and for comparison we have included the \citet{Sutherland_1993} cooling function for low metallicity plasma, Eq. (\ref{eq:SD_cooling}). The linear regression does indeed show a relationship albeit a weak one, with an exponent of $-0.28$. For the \citet{Sutherland_1993} cooling function we have taken the magnitude of cooling to be that at the minimum, Eq.~ (\ref{eq:SD_min}). The fitted $\beta$ calculated using this figure is a factor of $\sim 25\%$ lower than that using our Heaviside cooling function using the same normalisation. This is not quite as strong as the dependence suggested by Eq.~(\ref{eq:betamax}), of $-6/11 \approx -0.54$.

In Fig.~\ref{fig:sd_comparison} we make a further comparison between the \citet{Sutherland_1993} cooling function and our flat cooling function. We chose the run with the nearest normalisation ($\Lambda=1.5 \times 10^{-24} \, \rm erg \, cm^3 \, s^{-1}$) to that of the minimum of the \citet{Sutherland_1993} cooling function (Eq.~\ref{eq:SD_min}). We see a very similar phase distribution of the ISM, suggesting that the detailed structure of the cooling function above $10^4\, \rm K$ does not play a large role in determining the features of the ISM.

\begin{figure}
\centering
\includegraphics[width=\columnwidth]{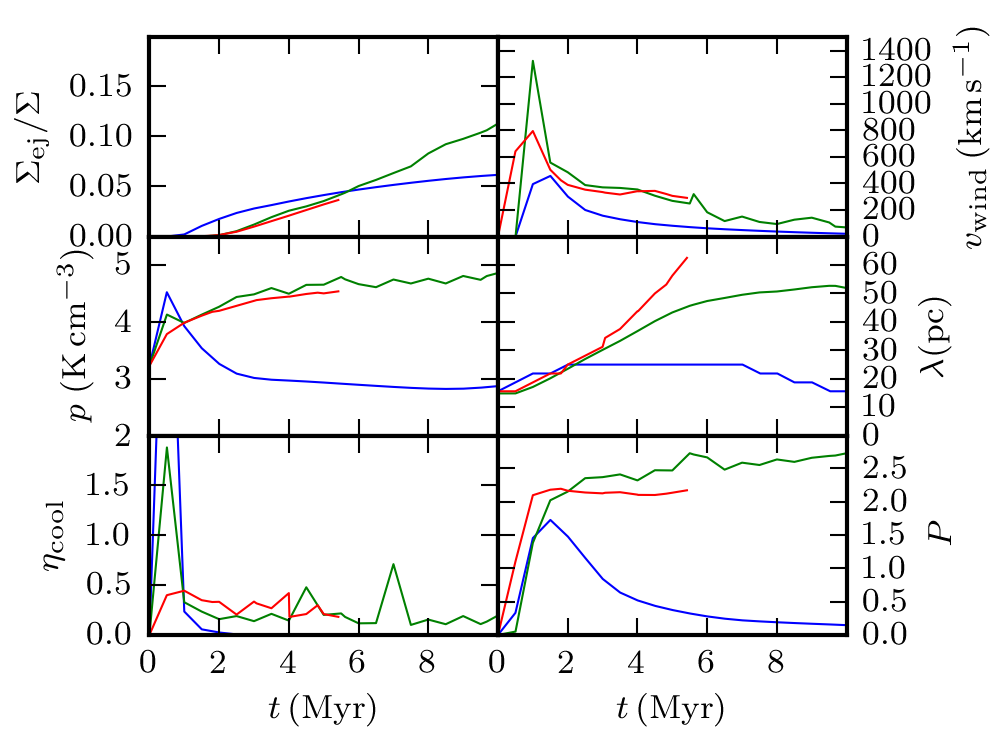}
\caption{Evolution of the simulations as a function of supernovae granularity. \emph{Green line} shows a run with $10^{51}\, \rm ergs$ per SN. \emph{Red line} is the same simulation, but with $50 \times$ the frequency of SNe, each releasing $1/50$th of the energy ($2\times 10^{49}$ ergs). \emph{Blue line} has SNe at $1/50$th of the frequency, with $50 \times$ the energy ($5\times 10^{52}$ ergs). }
\label{fig:sne_size_effect}
\end{figure}

In Fig.~\ref{fig:sne_size_effect} we have taken another high star formation rate simulation and adjusted the energy associated with a single supernova. Here, we keep the average rate of energy injected per unit time fixed, but inject the energy in more (less) frequent explosions with less (more) energy. The variation between these simulations is surprisingly large: 
the behaviour of the ISM is indeed quite sensitive to how smooth or stochastic the injected energy is.

\begin{figure}
\centering
\includegraphics[width=\columnwidth]{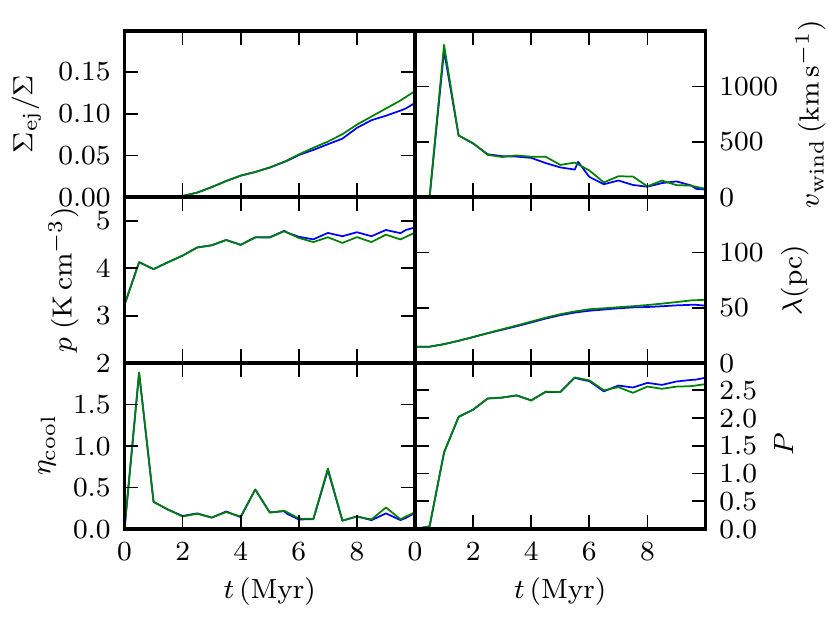}
\caption{As for Fig.\ref{fig:fiducial_convergence} but testing the effect of changing the vertical box size. \emph{Blue line} is the outflow estimated for a simulation with the fiducial box height (500 pc), \emph{green line} for 1 kpc.}
\label{fig:convergence_tallbox}
\end{figure}

\begin{figure}
\centering
\includegraphics[width=\columnwidth]{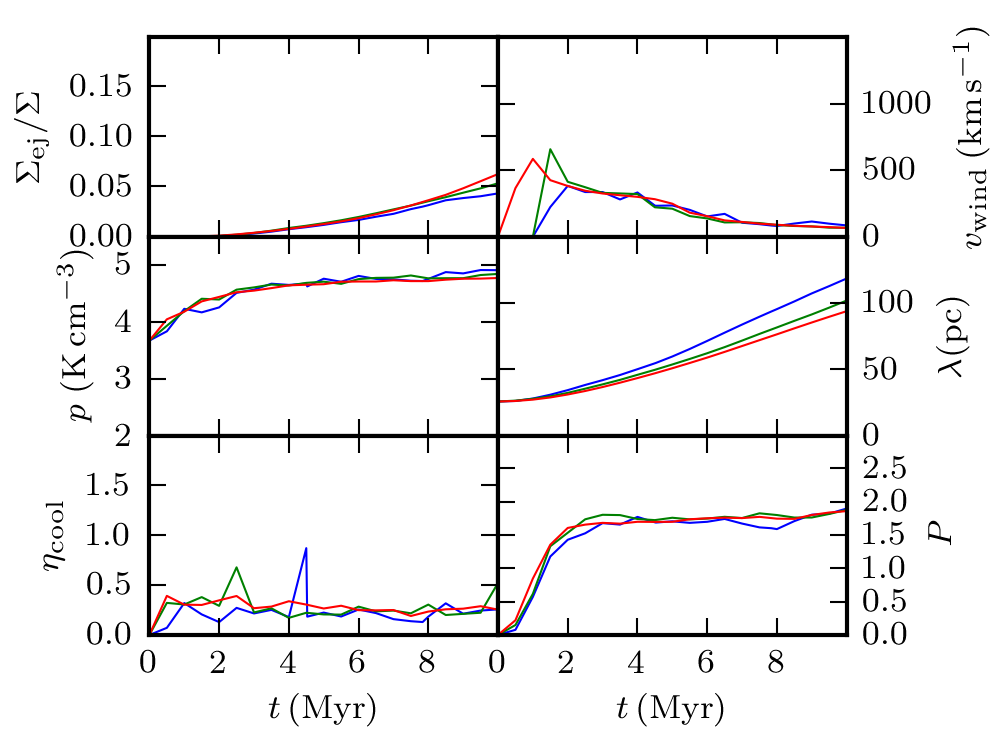}
\caption{As Fig.~\ref{fig:fiducial_convergence}, but for 3 different box sizes. \emph{Blue line} indicates the run at the L3 resolution, \emph{green line} with $2\times$ the box width and \emph{red line} with $4\times$ the box width.}
\label{fig:box_convergence}
\end{figure}

In Figs. \ref{fig:convergence_tallbox}-\ref{fig:box_convergence} we investigate the dependence of the simulations on the size of the simulation volume. In Fig. \ref{fig:convergence_tallbox} we adjust the  vertical size of the simulation volume, i.e. whether increasing the volume to simulate more of the outflow adjusts the dynamics, for example by allowing some material to fall back to the disk. All parameters are still computed for the original volume ($\pm 500$ pc), only the simulation volume has been expanded. All the parameters appear to be almost independent of this change. In Fig.~\ref{fig:box_convergence} we adjust the horizontal size of simulation volume, where we have multiplied the box width by a factors of $2$ and 4 respectively. The parameters here also show extremely good convergence, with the larger volumes generally showing less variation in values due to the reduced Poisson noise. The larger volumes also appear to show a marginal reduction in the evolution of the disk scale height.

\begin{figure*}
\vbox to220mm{\vfil 
\includegraphics[width=2\columnwidth]{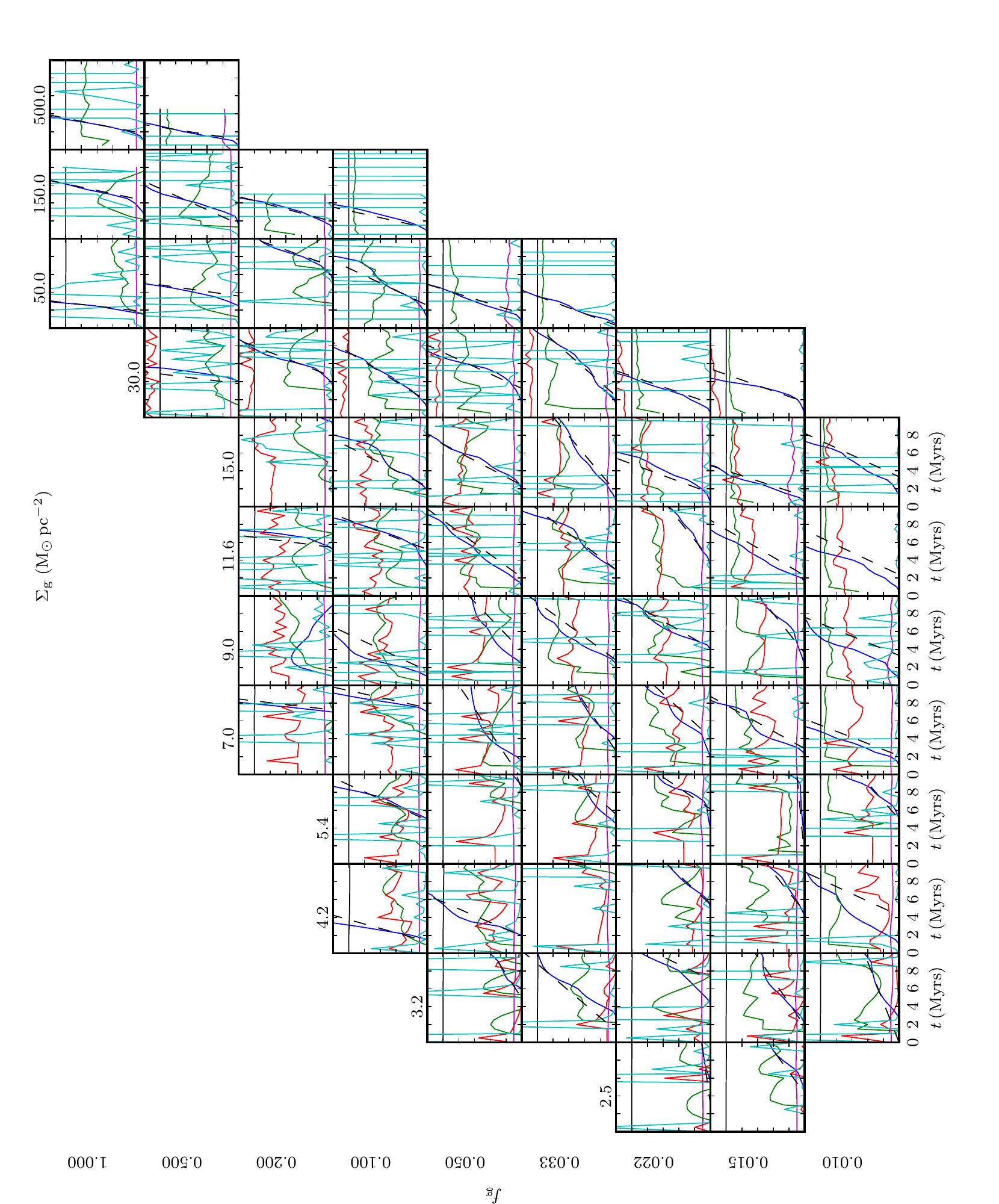}
\caption{As Fig.~\ref{fig:single_evolution}, but for all simulations varying $\Sigma_{\rm g}$ and $f_{\rm g}$ in Table~\ref{tab:parameters}. Additional \emph{black dashed lines} have been added to show a fit to ejection (\emph{solid blue lines}) given by Eq.~(\ref{eq:betafit}). $\Sigma_{\rm g1}\equiv \Sigma / 1 M_\odot \, \rm pc^{-1}$ runs from $1.5$ to $11.6$ from left to right panels. $f_{\rm g}$ runs from $0.01$ to $0.05$ from lower to upper panels. \emph{Blue line}, the amount of gas that has been ejected from the simulation, has now been scaled to units of $5\, \rm M_\odot \, pc^{-2}$ for clarity. \emph{Green line} shows $1+P$,  \emph{red line}, the mean pressure, \emph{black line}, the fraction of gas remaining in the simulation, \emph{magenta line}, the height of the disk and \emph{cyan line}, the very stochastic instantaneous cooling rate as a fraction of the mean SNe energy injection rate. }
\label{fig:sliceA}
\vfil}
\end{figure*}

Finally, in Fig. \ref{fig:sliceA} we have constructed equivalent graphs to that of Fig.~\ref{fig:single_evolution}, but now showing all simulations varying $\Sigma_{\rm g}$ and $f_{\rm g}$ in Table~\ref{tab:parameters}. Each panel shows the time evolution of a single simulation, showing the surface density, porosity, instantaneous cooling rate, disk height, mass ejected and pressure, along with a ramp function fit, Eq.~ (\ref{eq:betafit})), to the mass ejection rates.  We can see a strong evolution of the feedback from top left to lower right, i.e. at high surface densities and low gas fractions the simulations develop much stronger mass ejection rates and pressures, and the disk is more heavily disrupted. Note that the mass ejection rate has not been normalised by the surface density, so much of the increase is due to the increased star formation in the higher surface density disks. In a couple of the high surface density panels the simulation has failed early although there are enough data points to perform a fit to the mass ejection. Although there is considerable stochasticity in the 2 parameter fits, they seem quite robust.

In conclusion, these studies demonstrate that our simulations are effective at modelling a SN driven ISM and resilient to changes in numerical parameters. The exact nature of the cooling function exhibits little effect on the disk evolution, in fact the limiting factor is largely the physical granularity of the discrete SNe and their locations in the disk. To this end reducing the scatter in our disk property dependencies could be acheived by taking a larger ensemble of runs or alternatively by simulating larger disk areas, either of which increases the total number of SNe introduced.

\end{appendix}
\label{lastpage}
\end{document}